\documentclass[journal]{IEEEtran}
\usepackage{cite}
\usepackage{amsmath,amssymb,amsfonts}
\usepackage{algorithmic}
\usepackage{graphicx}
\usepackage{textcomp}
\usepackage{tikz}
\usepackage{todonotes}
\usepackage{eucal}

\usepackage{xcolor}
\usepackage{mathrsfs}
\usepackage{hyperref}
\hypersetup{
  pdfstartview=FitH,
  colorlinks=true,
}

\newtheorem{theorem}{Theorem}
\newtheorem{lemma}{Lemma}
\newtheorem{proposition}{Proposition}

\newtheorem{definition}{Definition}
\newtheorem{assumption}{Assumption}

\newtheorem{example}{Example}
\newtheorem{remark}{Remark}

\newcommand{\diff}{\mathrm{d}}
\newcommand{\e}{\mathrm{e}}
\newcommand{\trans}{\mathsf{T}}

\newcommand{\bbR}{\mathbb{R}}
\newcommand{\bbZ}{\mathbb{Z}}

\newcommand{\bbT}{\mathbb{T}}

\newcommand{\calC}{\mathcal{C}}
\newcommand{\calE}{\mathcal{E}}
\newcommand{\calG}{\mathcal{G}}

\newcommand{\calK}{\mathcal{K}}
\newcommand{\calL}{\mathcal{L}}

\newcommand{\calS}{\mathcal{S}}
\newcommand{\calT}{\mathcal{T}}

\newcommand{\calV}{\mathcal{V}}

\newcommand{\bbS}{\mathbb{S}}

\DeclareMathOperator{\diag}{diag}
\DeclareMathOperator{\Span}{span}

\DeclareMathOperator{\image}{im}
\DeclareMathOperator{\kernel}{ker}

\DeclareMathOperator{\dist}{dist}

\newcommand{\bbone}{\text{\usefont{U}{bbold}{m}{n}1}}

\begin{document}

\title{Cluster Synchronization of Kuramoto Oscillators and the Method of Averaging}
\author{Rui Kato and Hideaki Ishii, \textit{Fellow, IEEE}%
\thanks{This work was supported in the part by JSPS under Grants-in-Aid for Scientific Research (B) Grant No.~JP18H01460 and for JSPS Research Fellow Grant No.~JP21J21262, and by JST-Mirai program Grant No. 18077648.}%
\thanks{R. Kato and H. Ishii are with the Department of Computer Science, Tokyo Institute of Technology, Yokohama, 226-8502, Japan. (e-mail: kato@sc.dis.titech.ac.jp; ishii@c.titech.ac.jp).}}
\maketitle

\begin{abstract}
  Rigorous conditions for cluster synchronization of Kuramoto oscillators are presented.
  The method of averaging plays an important role in stability analysis, but the standard Lyapunov's second method is not applicable due to the lack of uniform continuity.
  This paper contributes to overcoming this difficulty with the help of nonmonotonic Lyapunov functions.
  Our extensions of averaging in stability theory are key to derive the two interrelated cluster synchronization conditions: (i) the coupling strengths between clusters are sufficiently weak and/or (ii) the natural frequencies are largely different between clusters.
  Cluster phase cohesiveness in the absence of network partitions ensuring the existence of invariant manifolds is also investigated.
  Moreover, we apply our theoretical findings to brain networks and exhibit certain relations among network parameters and functional connectivity.
\end{abstract}


\section{Introduction}

Synchronization in complex networks has attracted interests over the past several decades (see, e.g., \cite{Arenas_PhysRep2008} and the references therein).
Application areas include power systems \cite{Motter_NatPhys2013}, networked robotics \cite{Francis_book2016}, physiological rhythms \cite{Glass_nature2001}, and neuronal oscillations \cite{Buzsaki_science2004}, to name a few.
The phenomenon that all physical quantities are consistent among the whole network is called \emph{full synchronization}, whereas the phenomenon that the quantities are matched within some groups is called \emph{cluster synchronization} \cite{Belykh_Chaos2008,Lu_Chaos2010}.
Although we do not consider in this paper, there are also other types of synchronization such as \emph{remote synchronization} \cite{Nicosia_PRL2013} and \emph{chimera states} \cite{Cho_PRL2017}.

Compared with full synchronization, the mechanisms of cluster synchronization have not been much explored in spite of its scientific importance.
Representative examples of cluster synchronization are functional connectivity in brain networks \cite{Stam_2012} and group consensus in opinion dynamics \cite{Hegselmann_2002}.
Recently, brain functional connectivity has been examined in relation with neuronal diseases \cite{Bassett_2009}.
On the other hand, stability analysis and control design for cluster synchronization have been studied from various perspectives \cite{Pham2007,Pogromsky_2008,Wu2009,Lu_PLA2009,Liu2011,Xia_2011}.
The objective of this paper is to clarify some stability aspects in cluster synchronization phenomena.
In particular, we consider coupled phase oscillators described by the Kuramoto model \cite{Kuramoto_book1984}.
In the following, we briefly review recent results related to this topic, and then, we explain the main tools used in the theoretical analysis.

\subsection{Synchronization Conditions for Kuramoto Oscillators}

The Kuramoto model is one of the most well known mathematical models of oscillators.
In spite of its simple representation, the Kuramoto model is useful in many application areas \cite{Acebron_RMP2005}.
Furthermore, dynamical properties such as stability and synchronizability have been studied for a long time.
In particular, graph-theoretic characterizations of Kuramoto oscillators and the basic synchronization conditions appeared in the seminal paper \cite{Jadbabaie_ACC2004}.
Specifically, the Kuramoto oscillators achieve frequency synchronization if the coupling strengths are sufficiently strong compared with the natural frequency heterogeneity.
Many other results can be found in the recent surveys \cite{Dorfler_Automatica2014,Wu-Li_2020}.

On the other hand, cluster synchronization of Kuramoto oscillators was recently considered in \cite{Schaub_Chaos2016,Menara_TCNS2020}, and \textcolor{black}{this paper mainly follows their problem settings.}
From \cite{Menara_TCNS2020}, it is known that cluster synchronization can be achieved if (i) the coupling strengths within clusters are sufficiently stronger than those between clusters and/or (ii) the natural frequency differences between clusters tend to infinity.
Moreover, approximated but tighter stability conditions were derived in \cite{Menara_ACC2019} based on the frequency-domain analysis and the small-gain theorem.
A slightly different setting was considered in \cite{Qin_TAC2021}, where phase cohesiveness within clusters was addressed instead of exact cluster synchronization.
To explore exact cluster synchronization, we need a considerable restriction in the cluster structure.
If the underlying network is partitioned by a so-called external equitable partition, then the cluster synchronization manifold is an invariant manifold \cite{Schaub_Chaos2016} (see also \cite{Tiberi_CDC2017}).
This restriction was relaxed in \cite{Menara_CDC2019}, where applications to brain networks were also considered.


\subsection{The Method of Averaging in Stability Theory}

When we analyze stability properties of Kuramoto oscillators, averaging can be applied to the fast oscillations due to the inter-cluster natural frequency differences.
We will extend the method of averaging in stability theory, which will play fundamental roles in the analysis of cluster synchronization of Kuramoto oscillators.

The method of averaging is useful in approximation of solutions to periodic or almost periodic systems \cite{Sanders-Verhulst-Murdock_book2007} as well as in stability analysis of those systems \cite{Hapaev_book1993}.
By the averaging principle, the average dynamics approximates the original dynamics on some finite interval.
In general, the approximation level and the time interval on which the approximation is valid depend on the value of the selected perturbation parameter.
Under an additional assumption, one can extend the time interval on which the approximation is valid to the interval of infinite length.
In such a case, one can construct a Lyapunov function based on the average system and can show that it is also a Lyapunov function for the original system (see, e.g., \cite{Hale_book1980,Khalil_book2002}).
However, this additional assumption is somewhat restrictive since it requires that the right-hand side of the differential equation is continuous with respect to the perturbation parameter uniformly in time.

One aim of this paper is to relax this requirement based on the ideas in \cite{Aeyels_TAC1998,Aeyels_TAC1999}.
Particularly, we rely on the method of nonmonotonic Lyapunov functions \cite{Michel-Hou-Liu_book2015} instead of the standard Lyapunov's second method.
A nonmonotonic Lyapunov function does not require that its time derivative along any solution of the underlying equation is negative definite.
Instead, it is required that the value of the nonmonotonic Lyapunov function must decrease over some discrete set from the time interval.

\subsection{Contributions and Paper Structure}

In this paper we derive rigorous cluster synchronization conditions for Kuramoto oscillators.
Compared with the previous works mentioned above, we regard the Kuramoto model as a coupled system of the slow intra-cluster subsystem and the fast inter-cluster subsystem.
As a result, the problem is translated into the partial stability problem for an ordinary differential equation in the singular perturbation form.
More precisely, we consider stability of a ``partial equilibrium'' of the system (see \cite{Vorotnikov_2005} for various definitions of partial stability).
Introducing a perturbation parameter defined by the smallest inter-cluster natural frequency difference, we apply averaging techniques to obtain stability conditions.
The contributions of this paper are summarized as follows:

\begin{enumerate}
  \item We provide some stability and boundedness criteria for certain classes of nonautonomous systems based on the method of averaging. Our approach does not require uniform continuity of the perturbation parameter for the system under consideration. The obtained results also play fundamental roles in the investigations of cluster synchronization of Kuramoto oscillators. (Section~\ref{sec:averaging})
  \item We provide new stability conditions for cluster synchronization of Kuramoto oscillators for general cases and for special cases with two clusters. In contrast to \cite{Menara_TCNS2020}, our approach is unified in the sense that an interrelation between the two key factors for cluster synchronization can be examined. (Sections~\ref{subsec:main-A} and \ref{subsec:main-B})
  \item We characterize cluster phase cohesiveness for the Kuramoto model in terms of the coupling strengths and the natural frequencies. This result is applicable to the case where the cluster synchronization manifold is not an invariant set. (Section~\ref{subsec:main-C})
  \item We apply our theoretical findings to the brain networks constructed from empirical data. By numerical simulation, we visualize certain relations among the coupling strengths, the natural frequencies, and the functional connectivity. This analysis indicates that brain functional connectivity depends on network parameters. (Section~\ref{sec:brain})
\end{enumerate}

This paper can be divided into three parts.
In the first part, we describe conditions for cluster synchronization and cluster phase cohesiveness of Kuramoto oscillators in Sections~\ref{sec:preliminaries} and \ref{sec:main}.
The second part is related to some extensions of the method of averaging in stability theory and is placed at Section~\ref{sec:averaging}.
Finally, the third part is devoted to demonstrate the theoretical findings through the case study from network neuroscience in Section~\ref{sec:brain}.
A preliminary version of this paper appeared as the conference paper \cite{Kato_ECC2021}; in the current work, we provide further extensions and the proofs of all results along with more extensive simulations.

\subsection{Notations}

Let $\bbR$, $\bbR_+$, $\bbZ$, and $\bbZ_+$ be the set of real numbers, nonnegative real numbers, integers, and nonnegative integers, respectively.
Let $\bbS^1$ be the unit circle and $\bbT^n = \bbS^1 \times \cdots \times \bbS^1$ the $n$-torus.
In this paper, we identify $\bbS^1$ as the group $\bbR/(2 \pi \bbZ)$ with addition modulo $2 \pi$, that is, $\theta \in \bbS^1$ implies $\theta + 2 \pi n \in \bbS^1$ for each $n \in \bbZ$.
With this notation, any phase difference $\theta - \tilde{\theta}$ again belongs to $\bbS^1$, where $\theta,\tilde{\theta} \in \bbS^1$.
The $n$-vector whose entries are all $1$ is denoted by $\bbone_n$.
The imaginary unit is denoted by $j$ (the same symbol is also used as an index, but which meaning is used is clear).
A function $\alpha \colon [0,\infty) \to [0,\infty)$ is of class $\calK$ if it is continuous, strictly increasing, and $\alpha(0) = 0$.
A function $\beta \colon [0,\infty) \to [0,\infty)$ is of class $\calL$ if it is continuous, nonincreasing, and $\beta(s) \to 0$ as $s \to \infty$.
We define the distance from a point $x \in \bbR^n$ to a subset $S \subset \bbR^n$ by $\dist(x,S) := \inf_{y \in S} \|x - y\|$.

\section{Preliminaries} \label{sec:preliminaries}

Here, we formulate the problem on cluster synchronization of Kuramoto oscillators \cite{Schaub_Chaos2016,Menara_TCNS2020}.
For the sake of subsequent developments, we transform the Kuramoto model into differential equations in terms of phase differences.
We then associate the problem with partial stability and explain the main difficulty.

\subsection{Problem Formulation} \label{subsec:problem_formulation}

Consider a connected undirected graph $\calG = (\calV,\calE)$, where $\calV = \{1,\ldots,N\}$ is the set of nodes and $\calE \subset \calV \times \calV$ is the set of edges.
Let $A = [a_{ij}] \in \bbR^{N \times N}$ be the weighted adjacency matrix of $\calG$, where $a_{ij} > 0$ if $(i,j) \in \calE$ and $a_{ij} = 0$ otherwise.
After selecting an enumeration and an orientation for all (undirected) edges, we define the oriented incidence matrix $B = [b_{ie}] \in \bbR^{N \times |\calE|}$, where
\begin{align*}
  b_{ie} =
  \begin{cases}
    1  & \text{if node $i$ is the sink of edge $e$}, \\
    -1 & \text{if node $i$ is the source of edge $e$}, \\
    0  & \text{otherwise}.
  \end{cases}
\end{align*}
Associated with the enumeration selected above, we define the weight matrix by $W = \diag[(a_{ij})_{(i,j) \in \calE}]$.

We introduce a graph partition \cite{Godsil-Royle_book2001}.
For an integer $r \ge 2$, let $\Pi = \{\calC_1,\ldots,\calC_r\}$ be a \emph{nontrivial partition} of $\calV$, i.e., the following conditions hold: (i) $\calC_p \ne \emptyset$ for all $p \in \{1,\ldots,r\}$, (ii) $\calC_p \cap \calC_q = \emptyset$ if $p \ne q$, and (iii) $\bigcup_{p=1}^r \calC_p = \calV$.
We call $\calC_1,\ldots,\calC_r$ \emph{clusters} of the graph $\calG$.
The graph $\calG$ can now be decomposed as $\calG = \bigcup_{p=1}^r \calG_p \cup \calG_\mathrm{inter}$, where $\calG_p = (\calC_p,\calE_p)$ and $\calG_\mathrm{inter} = (\calV,\calE_\mathrm{inter})$.
Here, we have defined $\calE_p := \{(i,j) \in \calE : i,j \in \calC_p\}$ and $\calE_\mathrm{inter} := \calE \setminus \bigcup_{p=1}^r \calE_p$.
In this paper, it is assumed that the subgraph $\calG_p$ is connected for every $p \in \{1,\ldots,r\}$\footnote{This assumption is removed in \cite{Menara_CSL2022} with the help of relay nodes.}.
Finally, we assume without loss of generality that each cluster contains at least two nodes.

We consider Kuramoto oscillators coupled on the graph $\calG$, where each node represents an oscillator and each edge represents an interconnection of two oscillators.
The $i$th oscillator in the Kuramoto model is governed by
\begin{align*}
  \dot\theta_i = \omega_i + \sum_{j=1}^N a_{ij} \sin(\theta_j - \theta_i), \quad i \in \{1,\ldots,N\},
\end{align*}
where $\theta_i \colon \bbR_+ \to \bbS^1$ is the phase and $\omega_i \in \bbR$ is the natural frequency.
The equations above can be written in vector form:
\begin{align}
  \dot\theta = \omega - B W \sin(B^\trans \theta), \label{eq:Kuramoto_model_vector}
\end{align}
where $\theta = [\theta_1\;\cdots\;\theta_N]^\trans$ and $\omega = [\omega_1\;\cdots\;\omega_N]^\trans$.

If a solution $\theta(\cdot)$ of \eqref{eq:Kuramoto_model_vector} satisfies $\dist(\theta(t),\Span{(\bbone_N)}) \to 0$ as $t \to \infty$, then the oscillators are said to have achieved full synchronization.
Note that this is possible only if $\omega \in \Span(\bbone_N)$, that is, all natural frequencies are identical.
In this paper, we are interested in cluster synchronization, which is formally defined below.

The \emph{cluster synchronization manifold} with respect to the partition $\Pi = \{\calC_1,\ldots,\calC_r\}$ is defined by
\begin{align*}
  \calS_\Pi := \{\theta \in \bbT^N : \theta_i = \theta_j,\ i,j \in \calC_p,\ p \in \{1,\ldots,r\}\}.
\end{align*}
If $\dist(\theta(t),\calS_\Pi) \to 0$ as $t \to \infty$, then the oscillators are said to have achieved cluster synchronization.
Now, we define the stability of $\calS_\Pi$ to be investigated.

\begin{definition} \label{def:manifold_stability}
  The cluster synchronization manifold $\calS_\Pi$ is said to be \emph{(locally) exponentially stable} for \eqref{eq:Kuramoto_model_vector} if there exist $C \ge 1$, $\lambda > 0$, and $\delta > 0$ such that $\dist(\theta(0),\calS_\Pi) < \delta$ implies
  \begin{align*}
    \dist(\theta(t),\calS_\Pi) \le C \e^{-\lambda t} \dist(\theta(0),\calS_\Pi)
  \end{align*}
  for all $t \in \bbR_+$.
\end{definition}

Obviously, the above notion requires that the cluster synchronization manifold is forward invariant for \eqref{eq:Kuramoto_model_vector}, i.e., $\theta(0) \in \calS_\Pi$ implies $\theta(t) \in \calS_\Pi$ for all $t \in \bbR_+$.
It is known that this is ensured under the following two assumptions \cite{Schaub_Chaos2016}:

\begin{assumption} \label{ass:natural_frequency}
  It holds that $\omega_i = \omega_j$ for all $i,j \in \calC_p$ and all $p \in \{1,\ldots,r\}$.
\end{assumption}

\begin{assumption} \label{ass:external_equitable_partition}
  It holds that $\sum_{l \in \calC_q} (a_{il} - a_{jl}) = 0$ for all $i,j \in \calC_p$ and all $p,q \in \{1,\ldots,r\}$ with $p \ne q$.
\end{assumption}

Assumption~\ref{ass:natural_frequency} restricts the oscillators' natural frequencies in each cluster to be identical.
Assumption~\ref{ass:external_equitable_partition} means that any two nodes in the same cluster have equal weight sums with respect to each other cluster.
These two assumptions are also necessary for the forward invariance in certain cases (see \cite{Tiberi_CDC2017}).
The first assumption is a natural requirement whereas the second one is somewhat restrictive.
We will relax this aspect in Section~\ref{subsec:main-C}.

\begin{remark}
  A graph partition satisfying the condition in Assumption~\ref{ass:external_equitable_partition} is called an \emph{external equitable partition} (in short, EEP), which was introduced in \cite{Cardoso2007}.
  Such a partition is used in works of cluster synchronization with linear and nonlinear dynamics \cite{Gambuzza_Automatica2019}.
  The notion known as graph symmetry is also used in \cite{Pecora2014,Sorrentino2016}, and the relations between these two concepts are discussed in \cite{Schaub_Chaos2016}.
  We remark that, when the network size is large, there are a huge number of partitions that belong to the class of external equitable partitions.
  The trivial case is where each node is grouped into a single cluster.
  For algorithms to search external equitable partitions with minimal cardinality, we refer to \cite{O'Clery_PRE2013,Zhang_TAC2014}.
\end{remark}

\subsection{Reformulation as Partial Stability Problem}

To analyze synchronizability, it is useful to consider phase differences instead of phases themselves.
For this reason, we rewrite the Kuramoto model \eqref{eq:Kuramoto_model_vector} in terms of phase differences.
Then, we show that the problem of cluster synchronization is equivalent to the problem of partial stability.

By an appropriate enumeration of edges, we can write the matrices $B$ and $W$ as
\begin{align*}
  B =
  \begin{bmatrix}
    B_\mathrm{intra} & B_\mathrm{inter}
  \end{bmatrix}, \quad
  W =
  \begin{bmatrix}
    W_\mathrm{intra} & 0 \\
    0 & W_\mathrm{inter}
  \end{bmatrix},
\end{align*}
where $B_\mathrm{intra}$ and $W_\mathrm{intra}$ correspond to the subgraph $\bigcup_{p=1}^r \calG_p$ and $B_\mathrm{inter}$ and $W_\mathrm{inter}$ correspond to the subgraph $\calG_\mathrm{inter}$.

As with \cite{Menara_TCNS2020}, we consider a spanning tree $\calT = (\calV,\tilde{\calE})$ of the graph $\calG$, where $\tilde{\calE} \subset \calE$ is the set of selected edges.
Let $\tilde{B}$ be the incidence matrix of $\calT$.
Without loss of generality, we can write $\tilde{B} = [\tilde{B}_\mathrm{intra}\;\tilde{B}_\mathrm{inter}]$, where $\tilde{B}_\mathrm{intra}$ and $\tilde{B}_\mathrm{inter}$ are associated with the selected edges in $\bigcup_{p=1}^r \calE_p$ and $\calE_\mathrm{inter}$, respectively.
We denote by $n$ and $m$ the dimensions of the range spaces of $\tilde{B}_\mathrm{intra}^\trans$ and $\tilde{B}_\mathrm{inter}^\trans$, respectively.
Note that $n + m = N - 1$.

We can observe that there exists a matrix $R \in \bbR^{|\calE| \times (N-1)}$ such that $B^\trans = R \tilde{B}^\trans$.
Under an appropriate partition of $R$, we have the relation
\begin{align*}
  \begin{bmatrix}
    B_\mathrm{intra}^\trans \\
    B_\mathrm{inter}^\trans
  \end{bmatrix}
  = R
  \begin{bmatrix}
    \tilde{B}_\mathrm{intra}^\trans \\
    \tilde{B}_\mathrm{inter}^\trans
  \end{bmatrix}
  =
  \begin{bmatrix}
    R_1 & R_2 \\
    R_3 & R_4
  \end{bmatrix}
  \begin{bmatrix}
    \tilde{B}_\mathrm{intra}^\trans \\
    \tilde{B}_\mathrm{inter}^\trans
  \end{bmatrix}.
\end{align*}
The four submatrices of $R$ are given as follows \textcolor{black}{(see Appendix~\ref{app:derivation} for their derivations)}:
\begin{alignat*}{2}
  R_1 &:= B_\mathrm{intra}^\trans (\tilde{B}_\mathrm{intra}^\trans Q)^\dagger, & \quad R_2 &:= 0, \\
  R_3 &:= B_\mathrm{inter}^\trans (\tilde{B}_\mathrm{intra}^\trans Q)^\dagger, & \quad R_4 &:= B_\mathrm{inter}^\trans (\tilde{B}_\mathrm{inter}^\trans P)^\dagger,
\end{alignat*}
where $P := I_N - \tilde{B}_\mathrm{intra} \tilde{B}_\mathrm{intra}^\dagger$ and $Q := I_N - \tilde{B}_\mathrm{inter} \tilde{B}_\mathrm{inter}^\dagger$.

For the Kuramoto model \eqref{eq:Kuramoto_model_vector}, we define the two vectors $x := \tilde{B}_\mathrm{intra}^\trans \theta$ and $z := \tilde{B}_\mathrm{inter}^\trans \theta$, which represent the intra- and inter-cluster phase differences, respectively.
\textcolor{black}{Notice that $x(t) \in \bbT^n$ and $z(t) \in \bbT^m$ for all $t \in \bbR_+$.}
Then, we set a perturbation parameter $\varepsilon$ as the reciprocal of the smallest inter-cluster natural frequency difference.
Without loss of generality, we assume $\varepsilon$ to be positive, namely,
\begin{align}
  \varepsilon := \frac{1}{\min_{(i,j) \in \calE_\mathrm{inter}} |\omega_i - \omega_j|}. \label{eq:epsilon}
\end{align}

Under Assumption~\ref{ass:natural_frequency}, we obtain the system of differential equations in the singular perturbation form:
\begin{align}
  \dot{x} &= f(x,z),                                \label{eq:intra-cluster} \\
  \varepsilon \dot{z} &= \eta + \varepsilon g(x,z), \label{eq:inter-cluster}
\end{align}
where
\begin{align}
  f(x,z) &:= \Gamma_1 \sin(R_1 x) + \Gamma_2 \sin(R_3 x + R_4 z), \label{eq:def_f} \\
  g(x,z) &:= \Gamma_3 \sin(R_1 x) + \Gamma_4 \sin(R_3 x + R_4 z). \label{eq:def_g}
\end{align}
The four coefficient matrices are defined as follows:
\begin{alignat*}{2}
  \Gamma_1 &:= -\tilde{B}_\mathrm{intra}^\trans B_\mathrm{intra} W_\mathrm{intra}, & \quad \Gamma_2 &:= - \tilde{B}_\mathrm{intra}^\trans B_\mathrm{inter} W_\mathrm{inter}, \\
  \Gamma_3 &:= -\tilde{B}_\mathrm{inter}^\trans B_\mathrm{intra} W_\mathrm{intra}, & \quad \Gamma_4 &:= - \tilde{B}_\mathrm{inter}^\trans B_\mathrm{inter} W_\mathrm{inter}.
\end{alignat*}
The vector $\eta = \varepsilon \tilde{B}_\mathrm{inter}^\trans \omega$ represents the ratio of the natural frequency differences with respect to $\varepsilon$.
We note that the relation between $\omega$ and $\varepsilon$ is nonlinear and nonsmooth as in \eqref{eq:epsilon} and so is the dependence of $\eta$ on $\varepsilon$.
However, this relation does not affect the results since our results are independent of the specific values of $\eta$.

As discussed earlier, the cluster synchronization manifold $\calS_\Pi$ is forward invariant under Assumptions~\ref{ass:natural_frequency} and \ref{ass:external_equitable_partition}, and hence, we have $f(0,z) \equiv 0$.
Thus, stability of the partial equilibrium $x = 0$ is equivalent to that of the cluster synchronization manifold $\calS_\Pi$.
Specifically, the stability in Definition~\ref{def:manifold_stability} is compatible with the partial stability defined below \cite[Chap.~4]{Haddad-Chellaboina_book2008}.

\begin{definition} \label{def:partial_stability}
  \textcolor{black}{The partial equilibrium $x = 0$ of \eqref{eq:intra-cluster} and \eqref{eq:inter-cluster} is said to be \emph{exponentially stable uniformly in $z$} if there exist $C \ge 1$, $\lambda > 0$, and $\delta > 0$ such that $\|x(0)\| < \delta$ implies $\|x(t)\| \le C \e^{-\lambda t} \|x(0)\|$ for all $t \in \bbR_+$ uniformly in the initial states $z(0)$ of \eqref{eq:inter-cluster}.}
\end{definition}

\textcolor{black}{Note that the stability introduced above is a local concept. This allows us to describe the subsequent stability arguments without any consideration of the global structure of the $n$-torus $\bbT^n$, in which the $x$-variable resides. On the other hand, we are not concerned with any stability property of the $z$-variable.}

\begin{remark}
  A similar setting was considered in \cite{Qin_TAC2021b}, where the fast variable of a singularly perturbed system is restricted to be scalar.
  In that paper, by interpreting the fast variable as a new time variable, (partial) stability of a periodic system was analyzed through the method of averaging.
  In contrast, the equation \eqref{eq:inter-cluster} may not be a scalar system, and moreover, the system under consideration is not periodic in general.
  Thus, the results in \cite{Qin_TAC2021b} are not applicable in the current case.
  Another related work includes the so-called two-timescale averaging \cite{Teel2003}.
  However, in the current case, stability of the fast system is not considered.
\end{remark}

\subsection{Partial Linear Approximation} \label{subsec:partial_linearization}

Let us introduce the new time variable denoted by $\tau = t/\varepsilon$.
In this time scale the equations \eqref{eq:intra-cluster} and \eqref{eq:inter-cluster} are rewritten as
\begin{align}
  \frac{\diff x}{\diff \tau} &= \varepsilon f(x,z), \label{eq:dxdtau} \\
  \frac{\diff z}{\diff \tau} &= \eta + \varepsilon g(x,z). \label{eq:dzdtau}
\end{align}
Given initial states $x_0$ and $z_0$, we denote by $x = \xi(\tau,x_0,z_0,\varepsilon)$ and $z = \zeta(\tau,x_0,z_0,\varepsilon)$ the solutions to \eqref{eq:dxdtau} and \eqref{eq:dzdtau}, respectively.
Note that if $\varepsilon = 0$, then the solution to \eqref{eq:dzdtau} is given by $\zeta(\tau,x_0,z_0,0) = z_0 + \eta \tau$ and is independent of the variable $x$.

Recall that the problem under consideration is to determine partial stability with respect to $x$.
To tackle this problem, we employ partial linear approximation methods as considered in \cite{Miroshnik_2002,Hancock_2014}.
Since $f(0,z) \equiv 0$, the partial linear approximation of $f(x,z)$ with respect to $x$ is given by
\begin{align*}
  f(x,z) = \frac{\partial f}{\partial x}(0,z) x + o_x(x,z),
\end{align*}
where $\|o_x(x,z)\|/\|x\| \to 0$ as $x \to 0$ uniformly in $z$.
Hence, the system under consideration can be well approximated by
\begin{align}
  \frac{\diff x}{\diff \tau} &= \varepsilon \frac{\partial f}{\partial x}(0,z) x, \label{eq:linearized_x} \\
  \frac{\diff z}{\diff \tau} &= \eta + \varepsilon g(0,z). \label{eq:linearized_z}
\end{align}
If this system is exponentially stable with respect to $x$ uniformly in $z$, then the partial equilibrium $x = 0$ of \eqref{eq:dxdtau} and \eqref{eq:dzdtau} is (locally) exponentially stable uniformly in $z$.

Now, we explain how to analyze the partial stability of the linearized model in \eqref{eq:linearized_x} and \eqref{eq:linearized_z}, \textcolor{black}{where $f$ and $g$ are defined in \eqref{eq:def_f} and \eqref{eq:def_g}, respectively.}
Because the second equation is independent of the first one, it can be solved for a given initial state.
Substituting the solution $z = \zeta(\tau,0,z_0,\varepsilon)$ into the first equation, we obtain the linear time-varying system
\begin{align}
  \frac{\diff x}{\diff \tau} = \varepsilon J(\tau,\varepsilon,z_0) x, \label{eq:linearization}
\end{align}
where
\begin{align}
  J(\tau,\varepsilon,z_0) := \Gamma_1 R_1 + \Gamma_2 \diag[\cos(R_4 \zeta(\tau,0,z_0,\varepsilon))] R_3. \label{eq:def_J}
\end{align}
The partial stability problem can be solved by showing that any solution to \eqref{eq:linearization} starting at time $t = 0$ converges exponentially to zero uniformly in $z_0$.

\begin{remark} \label{rem:difficulty}
  Here, we explain the main difficulty of the problem under study.
  Clearly, we need to analyze the time-varying dynamics as described in \eqref{eq:linearization}.
  Because the system \eqref{eq:linearization} is almost periodic in $\tau$, the method of averaging seems to be useful for its analysis.
  However, we must deal with the technical problem due to nonuniform continuity with respect to $\varepsilon$.
  \textcolor{black}{In more detail, the solution $z = \zeta(\tau,0,z_0,\varepsilon)$ is in general not continuous with respect to $\varepsilon$ uniformly in $\tau$ and the same holds for the Jacobian $J(\tau,\varepsilon,z_0)$.
  The lack of this uniform continuity prevents us to apply the standard Lyapunov's second method in combination with perturbation techniques.}
  Note that the results in \cite{Hale_book1980,Khalil_book2002} presuppose such uniform continuity, and hence, we cannot utilize their results.
  Our idea is to approximate the solutions on a finite interval by averaging and then to apply the method of nonmonotonic Lyapunov functions in \cite{Michel-Hou-Liu_book2015} for guaranteeing the asymptotic convergence.
\end{remark}

\section{Some Extensions of Averaging \\ in Stability Theory} \label{sec:averaging}

The objective of this section is to show some extended stability criteria involving the method of averaging.
Based on the use of a nonmonotonic Lyapunov function, we show stability and boundedness of solutions for certain classes of systems.
The results in this section lay the foundations for deriving the cluster synchronization conditions in Section~\ref{sec:main}.

\subsection{Nonmonotonic Lyapunov Functions}

We briefly introduce the method of nonmonotonic Lyapunov functions \cite{Michel-Hou-Liu_book2015}.
Consider a nonautonomous system
\begin{align}
  \dot{x} = f(t,x), \label{eq:ODE}
\end{align}
where $f \colon \bbR_+ \times \bbR^n \to \bbR^n$ is a continuous function.
Assume that $f$ satisfies some hypotheses for ensuring the existence and uniqueness of solutions.
Let $\phi(t,t_0,x_0)$ denote the solution to \eqref{eq:ODE} defined on $[t_0,\infty)$ starting from $x_0$ at time $t_0$.

The following proposition from \cite{Michel-Hou-Liu_book2015} will be used to prove the subsequent results in this section.
We only consider nonmonotonic Lyapunov functions which do not depend on time.
In what follows, we use the notation $\mathcal{B}_\delta := \{x \in \bbR^n : \|x\| < \delta\}$ for $\delta > 0$.

\begin{proposition} \label{prop:nonmonotonic}
  Consider the nonautonomous system \eqref{eq:ODE}.
  Suppose that there exist a continuous function $V \colon \bbR^n \to \bbR_+$ and constants $\Omega \ge 0$ and $\delta > 0$ such that the following conditions hold:
  \begin{enumerate}
    \item There exist class $\mathcal{K}$ functions $\alpha_1,\alpha_2 \colon \bbR_+ \to \bbR_+$ such that for all $x \in \bbR^n$,
    \begin{align*}
      \alpha_1(\|x\|) \le V(x) \le \alpha_2(\|x\|).
    \end{align*}
    \item For each $(t_0,x_0) \in \bbR_+ \times \mathcal{B}_\delta$, there exists an unbounded strictly increasing time sequence $(t_k)_{k \in \bbZ_+}$ without any accumulation point (except for $+ \infty$) such that
    \begin{enumerate}
      \item there exists a class $\mathcal{K}$ function $\alpha_3 \colon \bbR_+ \to \bbR_+$, independent of $(t_0,x_0)$, such that
      \begin{align*}
        &\frac{1}{t_{k+1} - t_k} [V(\phi(t_{k+1},t_0,x_0)) - V(\phi(t_k,t_0,x_0))] \\
        &\quad{} \le - \alpha_3(\|\phi(t_k,t_0,x_0)\|)
      \end{align*}
      for all $k \in \bbZ_+$ for which $\|\phi(t_k,t_0,x_0)\| \ge \Omega$;
      \item there exists a continuous function $\psi \colon \bbR_+ \to \bbR_+$, independent of $(t_0,x_0)$, such that $\psi(0) = 0$ and
      \begin{align*}
        V(\phi(t,t_0,x_0)) \le \psi(V(\phi(t_k,t_0,x_0)))
      \end{align*}
      for all $t \in [t_k,t_{k+1}]$ and all $k \in \bbZ_+$.
    \end{enumerate}
  \end{enumerate}
  If $\Omega$ can be zero, then the zero solution is uniformly asymptotically stable.
  If $\Omega$ is positive, then the solutions starting in a small neighborhood of the origin are (locally) uniformly ultimately bounded.
\end{proposition}

\begin{remark}
  Note that the boundedness result in Proposition~\ref{prop:nonmonotonic} does not require that the system \eqref{eq:ODE} has an equilibrium point at the origin.
  Here, we let $\bar{\psi} := \max_{r \in [0,\alpha_2(\Omega)]} \psi(r)$.
  From the proof of Theorem~3.2.5 in \cite{Michel-Hou-Liu_book2015}, we know that under the conditions in the above proposition, for each $\alpha \in (0,\delta)$, there exists $T(\alpha) > 0$ such that $\|x_0\| < \alpha$ implies $\|\phi(t,t_0,x_0)\| \le \beta(\Omega)$ for all $t \ge t_0 + T(\alpha)$, where
  \begin{align*}
    \beta(\Omega) &:= \max \{\beta_1(\Omega),\alpha_1^{-1} \circ \beta_2(\Omega)\}, \\
    \beta_1(\Omega) &:= \max \{\bar{\psi},\alpha_1^{-1} \circ \alpha_2(\bar{\psi}),\alpha_1^{-1} \circ \alpha_2(\Omega)\}, \\
    \beta_2(\Omega) &:= \max \{\psi(r) : r \in [0,\alpha_2 \circ \beta_1(\Omega)]\}.
  \end{align*}
  Because $\bar{\psi} \to 0$ as $\Omega \to 0$ from the definitions, the ultimate bound $\beta(\Omega)$ of the solutions can be made small depending on the size of $\Omega$.
\end{remark}

\subsection{Asymptotic Stability of Linear Systems} \label{subsec:asymptotic_stability}

Consider the linear system described in the form
\begin{align}
  \dot{x} = \varepsilon A(t,\varepsilon) x, \label{eq:linear-averaging}
\end{align}
where $A \colon \bbR_+ \times (0,\varepsilon_0] \to \bbR^{n \times n}$ is a continuous and bounded function and $\varepsilon \in (0,\varepsilon_0]$ is a small parameter.
We impose the following two assumptions:
\begin{enumerate}
  \item There exists $A_\mathrm{av} \in \bbR^{n \times n}$ such that
  \begin{align}
    A_\mathrm{av} = \lim_{T \to \infty} \frac{1}{T} \int_t^{t + T} A(s,0) \,\diff s \label{eq:A_av}
  \end{align}
  holds uniformly in $t$.
  \item There exist a class $\calL$ function $\sigma \colon \bbR_+ \to \bbR_+$ and a class $\calK$ function $\rho \colon \bbR_+ \to \bbR_+$ such that for all $T > 0$ and all $\varepsilon \in (0,\varepsilon_0]$,
  \begin{align}
    \left\| \frac{1}{T} \int_t^{t + T} [A(s,\varepsilon) - A_\mathrm{av}] \,\diff s \right\| \le \sigma(T) + \rho(\varepsilon) \label{eq:integral}
  \end{align}
  holds uniformly in $t$.
\end{enumerate}

The following theorem extends some results in \cite{Aeyels_TAC1998,Aeyels_TAC1999} and can be established from Proposition~\ref{prop:nonmonotonic}.

\begin{theorem} \label{thm:linear_averaging}
  Suppose that the matrix $A_\mathrm{av}$ defined in \eqref{eq:A_av} is a Hurwitz matrix.
  Then, there exists $\varepsilon^* > 0$ such that if $\varepsilon < \varepsilon^*$, then the zero solution to \eqref{eq:linear-averaging} is exponentially stable.
\end{theorem}

\begin{IEEEproof}
  See Appendix~\ref{app:linear_averaging}.
\end{IEEEproof}

\begin{remark} \label{rem:averaging}
  We give some technical comments on Theorem~\ref{thm:linear_averaging} in comparison with existing results in the literature.
  Notice that the assumption in \eqref{eq:integral} is different from the one in \cite[Sec.~10.6]{Khalil_book2002} (see also \cite{Hale_book1980}).
  In particular, our condition characterizes the relations of the left-hand side with the length of the integral and with the perturbation parameter.
  It is compatible with the one in \cite{Khalil_book2002} by setting $\varepsilon = 0$.
  The main difference lies in the fact that we do not require $A(t,\varepsilon)$ in \eqref{eq:linear-averaging} to be continuous with respect to $\varepsilon$ uniformly in $t$.
  This relaxation is important in the study of cluster synchronization of Kuramoto oscillators as mentioned in Remark~\ref{rem:difficulty}.
  Moreover, in the previous works \cite{Aeyels_TAC1998,Aeyels_TAC1999}, only the cases where the right-hand side of the differential equation is independent of $\varepsilon$  were considered.
  Also, some related results can be found in \cite{Kosut1987} and \cite{Stilwell_SIADS2006}, where respectively, the authors considered the case where the uniform average does not exist and the case where the system is switched periodically.
\end{remark}

\subsection{Ultimate Boundedness of Nonlinear Systems} \label{subsec:ultimate_boundedness}

When the system does not have any equilibrium, we cannot apply the linearization principle for stability analysis.
To tackle such a problem, we analyze ultimate boundedness of solutions instead of asymptotic stability.

Consider the nonlinear system described in the form
\begin{align}
  \dot{x} = \varepsilon f(t,x,\varepsilon), \label{eq:nonlinear}
\end{align}
where $f \colon \bbR_+ \times D \times (0,\varepsilon_0] \to \bbR^n$ is a continuous function on an open domain $D$ containing the origin and $\varepsilon \in (0,\varepsilon_0]$ is a small parameter.
We impose the following assumptions, which are analogous to those in the previous subsection:
\begin{enumerate}
  \item There exists $f_\mathrm{av} \colon D \to \bbR^n$ such that for all $x \in D$,
  \begin{align*}
    f_\mathrm{av}(x) = \lim_{T \to \infty} \frac{1}{T} \int_t^{t + T} f(s,x,0) \,\diff s
  \end{align*}
  holds uniformly in $t$.
  Moreover, the average vector field satisfies $f_\mathrm{av}(0) = 0$ is Lipschitz continuous, i.e., there exists $L > 0$ such that $\|f_\mathrm{av}(x) - f_\mathrm{av}(y)\| \le L \|x - y\|$ for all $x,y \in D$.
  \item There exists $K > 0$ such that for all $x \in D$, all $T > 0$, and all $\varepsilon \in (0,\varepsilon_0]$,
  \begin{align*}
    \left\| \frac{1}{T} \int_t^{t + T} [f(s,x,\varepsilon) - f_\mathrm{av}(x)] \,\diff s \right\| \le K \left( \frac{1}{T} + \varepsilon \right)
  \end{align*}
  holds uniformly in $t$.
\end{enumerate}

The average system can be written as
\begin{align}
  \dot{y} = \varepsilon f_\mathrm{av}(y). \label{eq:nonlinear-average}
\end{align}
The following theorem indicates that the ultimate bound of the solutions can be made arbitrarily small by restricting $\varepsilon$ to be small.
A similar problem was considered in \cite{Teel-Peuteman-Aeyels_1999}, where the considered class of systems is more restrictive in the sense that the function $f$ is independent of $\varepsilon$.

\begin{theorem} \label{thm:nonlinear_averaging}
  Suppose that the zero solution to \eqref{eq:nonlinear-average} is exponentially stable.
  Then, for each $\beta > 0$, there exists $\varepsilon^* > 0$ such that if $\varepsilon < \varepsilon^*$, then corresponding to any $\alpha > 0$ sufficiently small, there exists $T(\alpha,\beta)$ such that for every $t_0 \in \bbR_+$,
  \begin{align*}
    \|x_0\| < \alpha \implies \|\phi(t,t_0,x_0)\| < \beta
  \end{align*}
  for all $t \ge t_0 + T(\alpha,\beta)$.
\end{theorem}

\begin{IEEEproof}
  See Appendix~\ref{app:nonlinear_averaging}.
\end{IEEEproof}

\section{Cluster Synchronization Conditions} \label{sec:main}

In this section we provide the full analysis for the cluster synchronization problem.
Based on the results in Section~\ref{sec:averaging}, we derive cluster synchronization conditions for Kuramoto oscillators.
The first two subsections are devoted to exact cluster synchronization under Assumptions~\ref{ass:natural_frequency} and \ref{ass:external_equitable_partition}.
In the last subsection, we will relax the invariance hypothesis guaranteed by Assumption~\ref{ass:external_equitable_partition} and consider cluster phase cohesiveness.

\subsection{Cluster Synchronization with EEPs: General Case} \label{subsec:main-A}

First, we do not restrict the number of clusters, i.e., $r$ can be arbitrary.
Recall that the problem is to study the linear time-varying system \eqref{eq:linearization} and to determine exponential convergence of its solutions.
To do so, we introduce the uniform average of the Jacobian at $\varepsilon = 0$:
\begin{align}
  J_\mathrm{av} := \lim_{T \to \infty} \frac{1}{T} \int_\tau^{\tau+T} J(s,0,z_0) \,\diff s. \label{eq:average-Jacobian}
\end{align}
By \eqref{eq:def_J}, we have $J_\mathrm{av} = \Gamma_1 R_1$.
We note that this is well defined because each entry of $\cos(\zeta(\tau,0,z_0,0))$ is periodic in $\tau$.
If these entries have a common period, then the vector-valued function $\cos(\zeta(\tau,0,z_0,0))$ is periodic in $\tau$.
Otherwise, it is almost periodic in $\tau$.

Observe that $J_\mathrm{av}$ can be written as the block diagonal matrix consisting of $J_1,\ldots,J_r$, where $J_p$ depends on the edge weights in the subgraph $\calG_p$.
Since $\calG_p$ is connected for each $p \in \{1,\ldots,r\}$, $J_\mathrm{av}$ is a Hurwitz matrix (see Lemma~3.1 in \cite{Menara_TCNS2020}).
We prove the following lemma to apply the result in Section~\ref{subsec:asymptotic_stability}.

\begin{lemma} \label{lem:integral-bound}
  There exists a constant $K > 0$ such that $K \to 0$ as $\|W_\mathrm{inter}\|_\infty \to 0$ and for all $T > 0$ and all $\varepsilon \in (0,\varepsilon_0]$,
  \begin{align*}
    \left\| \frac{1}{T} \int_\tau^{\tau + T} [J(\sigma,\varepsilon,z_0) - J_\mathrm{av}] \,\diff s \right\| \le K \left( \frac{1}{T} + \varepsilon \right)
  \end{align*}
  holds uniformly in $\tau$ and $z_0$.
\end{lemma}

\begin{IEEEproof}
  See Appendix~\ref{app:integral-bound}.
\end{IEEEproof}

The following theorem provides two interrelated cluster synchronization conditions.

\begin{theorem} \label{th:multi}
  Consider the Kuramoto model \eqref{eq:Kuramoto_model_vector} and a nontrivial graph partition $\Pi = \{\calC_1,\ldots,\calC_r\}$ such that the subgraphs $\calG_1,\ldots,\calG_r$ are all connected.
  Suppose that Assumptions~\ref{ass:natural_frequency} and \ref{ass:external_equitable_partition} hold.
  Then, the following statements are true:
  \begin{enumerate}
    \item There exists $a^* > 0$ such that if
    \begin{align*}
      a_{ij} < a^*, \quad (i,j) \in \mathcal{E}_\mathrm{inter},
    \end{align*}
    then $\calS_\Pi$ is exponentially stable.
    \item There exists $\omega^* > 0$ such that if
    \begin{align*}
      |\omega_i - \omega_j| > \omega^*, \quad (i,j) \in \mathcal{E}_\mathrm{inter},
    \end{align*}
    then $\calS_\Pi$ is exponentially stable.
  \end{enumerate}
\end{theorem}

\begin{IEEEproof}
  Following the proof of Theorem~\ref{thm:linear_averaging}, we define $\tau_k := kT$ for $k \in \bbZ_+$, where $T > 0$ is determined later.
  Let $\Phi(\tau_{k+1},\tau_k)$ be the state-transition matrix of \eqref{eq:linearization} on $[\tau_k,\tau_{k+1}]$.
  Also, the state-transition matrix on $[\tau_k,\tau_{k+1}]$ associated with the average Jacobian is given as $\Phi_\mathrm{av}(\tau_{k+1},\tau_k) = \e^{\varepsilon T J_\mathrm{av}}$.
  By the Peano--Baker series \cite{Rugh_book1996}, the difference $H(\tau_{k+1},\tau_k) = \Phi(\tau_{k+1},\tau_k) - \Phi_\mathrm{av}(\tau_{k+1},\tau_k)$ of the transition matrices satisfies
  \begin{align*}
    &\|H(\tau_{k+1},\tau_k)\| \\
    &\quad{} \le \varepsilon \left\| \int_{\tau_k}^{\tau_{k+1}} [J(\sigma,\varepsilon,z_0) - J_\mathrm{av}] \,\diff \sigma \right\| + 2 \sum_{l=2}^\infty \frac{(\alpha \varepsilon T)^l}{l!},
  \end{align*}
  where $\alpha := \sup \{\|J(\tau,\varepsilon,z_0)\| : \tau \in \bbR_+, \ \varepsilon > 0, \ z_0 \in \bbR^m\}$.
  Note from the definition that $\alpha$ is finite.
  Then, Lemma~\ref{lem:integral-bound} implies that
  \begin{align*}
    \|H(\tau_{k+1},\tau_k)\| \le K \varepsilon ( 1 + \varepsilon T)  + 2 (\e^{\alpha \varepsilon T} - 1 - \alpha \varepsilon T).
  \end{align*}

  We note that $J_\mathrm{av}$ has only the negative of the nonzero eigenvalues of the graph Laplacians for $\calG_1,\ldots,\calG_r$.
  This implies that $\|\Phi_\mathrm{av}(\tau_{k+1},\tau_k)\| \le \e^{- \lambda \varepsilon T}$, where $\lambda$ is the smallest algebraic connectivity of $\calG_1,\ldots,\calG_r$.
  It follows that
  \begin{align*}
    \|\Phi(\tau_{k+1},\tau_k)\| &\le \|\Phi_\mathrm{av}(\tau_{k+1},\tau_k)\| + \|H(\tau_{k+1},\tau_k)\| \\
    &\le \mu(K,\varepsilon,T),
  \end{align*}
  where
  \begin{align}
    \mu(K,\varepsilon,T) &:= \e^{- \lambda \varepsilon T} + K \varepsilon (1 + \varepsilon T) + 2 (\e^{\alpha \varepsilon T} - 1 - \alpha \varepsilon T). \label{eq:kappa}
  \end{align}
  If $\mu(K,\varepsilon,T) < 1$ for some $T > 0$, then $V(x) = \|x\|$ is a nonmonotonic Lyapunov function.
  By Theorem~\ref{thm:linear_averaging}, the solutions to \eqref{eq:linearization} converge exponentially to zero.

  Now, we show that for sufficiently small $K$ or $\varepsilon$, such a constant $T$ exists.
  If we fix $\varepsilon$, then we have $\mu(0,\varepsilon,T) = \e^{- \lambda \varepsilon T} + 2 (\e^{\alpha \varepsilon T} - 1 - \alpha \varepsilon T)$.
  It is clear that there exists $T > 0$ such that $\mu(0,\varepsilon,T) < 1$.
  This means by continuity that $\mu(K,\varepsilon,T) < 1$ is satisfied for sufficiently small $K > 0$, which corresponds to taking the edge weights between clusters to be small because of the property that $K \to 0$ as $\|W_\mathrm{inter}\|_\infty \to 0$.
  Next, we fix $K$ and explore the existence of $T$ such that $\mu(K,\varepsilon,T) < 1$ for sufficiently small $\varepsilon$.
  Since $J(\tau,\varepsilon,z_0)$ satisfies the hypotheses in Section~\ref{subsec:asymptotic_stability}, it is possible.
  By definition in \eqref{eq:epsilon}, the smaller $\varepsilon$ is, the larger the inter-cluster natural frequency differences are.
  To complete the proof, we notice that the above procedure is independent of $z_0$.
\end{IEEEproof}

\begin{figure}
  \centering
  \begin{tikzpicture}
    \node [rectangle, fill=red,   fill opacity=0.1, rounded corners, minimum width=170, minimum height=45] (C1) at (0, 1.6) {};
    \node [rectangle, fill=green, fill opacity=0.1, rounded corners, minimum width=170, minimum height=40] (C2) at (0,   0) {};
    \node [rectangle, fill=blue,  fill opacity=0.1, rounded corners, minimum width=170, minimum height=45] (C3) at (0,-1.6) {};
    \node [circle, draw=black, fill=white, minimum size=15] (v1) at (  -1, 1.8) {};
    \node [circle, draw=black, fill=white, minimum size=15] (v2) at (   1, 1.8) {};
    \node [circle, draw=black, fill=white, minimum size=15] (v3) at (  -2,   0) {};
    \node [circle, draw=black, fill=white, minimum size=15] (v4) at (   0,   0) {};
    \node [circle, draw=black, fill=white, minimum size=15] (v5) at (   2,   0) {};
    \node [circle, draw=black, fill=white, minimum size=15] (v6) at (-1.5,  -2) {};
    \node [circle, draw=black, fill=white, minimum size=15] (v7) at (   0,-1.2) {};
    \node [circle, draw=black, fill=white, minimum size=15] (v8) at ( 1.5,  -2) {};
    \draw (v1) -- (v2);
    \draw (v3) -- (v4);
    \draw (v4) -- (v5);
    \draw (v6) -- (v7);
    \draw (v7) -- (v8);
    \draw (v8) -- (v6);
    \draw (v1) -- (v3);
    \draw (v1) -- (v4);
    \draw (v2) -- (v4);
    \draw (v2) -- (v5);
    \draw (v3) -- (v6);
    \draw (v4) -- (v7);
    \draw (v5) -- (v8);
    \node [anchor=south west] (A) at (C1.south west) {\large $\mathcal{C}_1$};
    \node [anchor=south west] (B) at (C2.south west) {\large $\mathcal{C}_2$};
    \node [anchor=south west] (C) at (C3.south west) {\large $\mathcal{C}_3$};
  \end{tikzpicture}
  \caption{Network partitioned into three clusters}
  \label{fig:example}
\end{figure}
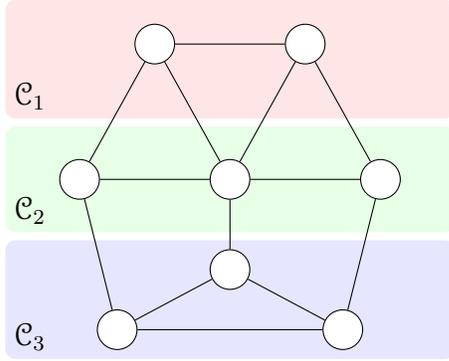

From Theorem~\ref{th:multi}, we can observe that cluster synchronization can be achieved if (i) the coupling strengths between clusters are sufficiently weak and/or (ii) the natural frequencies are largely different between clusters.
\textcolor{black}{In particular, we have provided a unified framework for stability analysis in that a certain relation between these two conditions are unveiled.
In fact, the inequality $\mu(K,\varepsilon,T) < 1$, where the definition is given in \eqref{eq:kappa}, is a unified sufficient condition for cluster synchronization.
Roughly speaking, for an appropriate choice of $T$, the smaller $K$ is or the smaller $\varepsilon$ is, the smaller $\mu(K,\varepsilon,T)$ is.
Note from the definition that if the intra-cluster coupling strengths are large, then the stability requirement becomes weak.
Due to its nonlinearity, it is difficult to solve this inequality explicitly, but we remark that a specific value of $K$ or $\varepsilon$ for which the stability condition is satisfied can be found numerically.
A more specific argument is visualized in the example given later.}

\textcolor{black}{Here, we highlight our approach in comparison with \cite{Menara_TCNS2020}.
A closely related result was presented in Theorem~3.5 in that paper, where the number of clusters is assumed to be two.
In that case, the evolution of the $z$-variable is periodic when $x = 0$, and thus, the analysis is relatively simple.
Our main contribution is to clarify that the above argument is also valid for the general case where the number of clusters is arbitrary by relying on the extended method of averaging developed in Section~\ref{sec:averaging}.
Moreover, Theorems~3.2 and 3.3 in \cite{Menara_TCNS2020} provide related cluster synchronization conditions for the multi-cluster case.
Compared with those results, we have shown that there is a finite threshold of the inter-cluster natural frequency differences above which cluster synchronization is achieved.
This is advantageous because Theorem~3.3 in \cite{Menara_TCNS2020} requires $|\omega_i - \omega_j|$ to be infinite for all $(i,j) \in \calE_\mathrm{inter}$.
The improvement of this point is also owing to the results in Section~\ref{sec:averaging}.}

\begin{remark}
  In \cite{Menara_ACC2019}, the authors developed an approximation method for deriving stability conditions involving both the coupling strengths and the natural frequencies.
  There, the problem is approximated properly and then a less conservative stability result is derived based on the small-gain theorem.
  Our result is different from \cite{Menara_ACC2019} in that the strict stability condition is provided though it is somewhat conservative.
  In the next subsection, we also provide a structural condition ensuring stability for some special cases, which can be less conservative.
\end{remark}

Here, we demonstrate the result of Theorem~\ref{th:multi} using a numerical example and visualize the interrelation between the two conditions just mentioned above.

\begin{figure}
  \centering
  \includegraphics[width=\columnwidth]{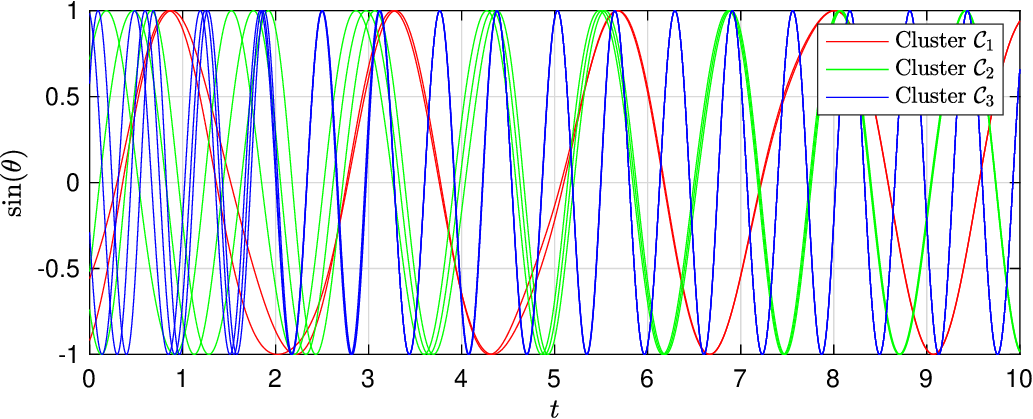}
  \caption{Phase evolutions of the oscillators in Example~\ref{ex:three_clusters}}
  \label{fig:sim_phases}
\end{figure}

\begin{figure}
  \centering
  \includegraphics[width=0.7\columnwidth]{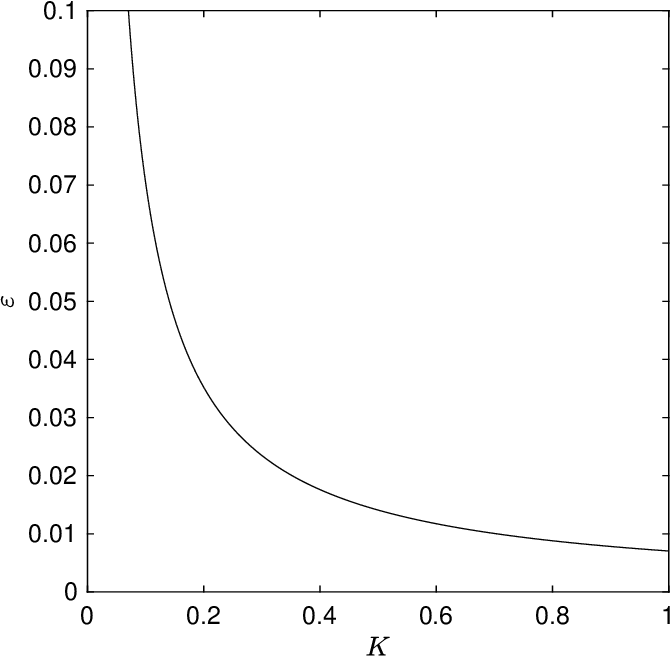}
  \caption{Curve for which there exists $T > 0$ such that $\mu(K,\varepsilon,T) = 1$ (the horizontal axis corresponds to the inter-cluster coupling strengths and the vertical axis corresponds to the inter-cluster natural frequency differences)}
  \label{fig:tradeoff_curve}
\end{figure}

\begin{example} \label{ex:three_clusters}
  Consider the network shown in Fig.~\ref{fig:example}, which is partitioned into three clusters as indicated.
  The adjacency matrix of this network is given by
  \begin{align*}
    A = \left[
    \begin{array}{cc|ccc|ccc}
      0 & a_1 & b_1 & b_1/2 & 0 & 0 & 0 & 0 \\
      a_1 & 0 & 0  & b_1/2 & b_1 & 0 & 0 & 0 \\ \hline
      b_1 & 0 & 0 & a_2 & 0 & b_2 & 0 & 0 \\
      b_1/2 & b_1/2 & a_2 & 0 & a_2 & 0 & b_2 & 0 \\
      0 & b_1 & 0 & a_2 & 0 & 0 & 0 & b_2 \\ \hline
      0 & 0 & b_2 & 0 & 0 & 0 & a_3 & a_3 \\
      0 & 0 & 0 & b_2 & 0 & a_3 & 0 & a_3 \\
      0 & 0 & 0 & 0 & b_2 & a_3 & a_3 & 0
    \end{array}
    \right].
  \end{align*}
  Note that Assumption~\ref{ass:external_equitable_partition} is satisfied for any parameter value.
  We set the edge weights as $a_1 = a_2 = a_3 = b_1 = b_2 = 0.5$.
  \textcolor{black}{The oscillators' natural frequencies are set as $\omega_i = 2.5$ for $i \in \calC_1$, $\omega_i = 5$ for $i \in \calC_2$, and $\omega_i = 10$ for $i \in \calC_3$.}
  We show the phase evolutions of the oscillators in Fig.~\ref{fig:sim_phases}.
  It can be observed that the oscillators are synchronized within clusters and oscillations of the inter-cluster phase differences appear.
  Furthermore, we present the curve for which there exists $T > 0$ such that $\mu(K,\varepsilon,T)$ in \eqref{eq:kappa} is equal to $1$ in Fig.~\ref{fig:tradeoff_curve}.
  This curve represents the specific values of the coupling strengths and the natural frequency differences for the stability conditions in Theorem~\ref{th:multi}.
  Stability is guaranteed for any pair $(K,\varepsilon)$ in the left-bottom area of the figure \textcolor{black}{(see also the discussion below Theorem~\ref{th:multi})}.
\end{example}

\subsection{Cluster Synchronization with EEPs: Special Case} \label{subsec:main-B}

The results in the previous subsection are somewhat conservative.
We here restrict ourselves to the case where the number of clusters is two, i.e., $r = 2$, and derive a possibly less conservative stability condition.

In the case of two clusters, the analysis is much simpler than the multi-cluster case (see also \cite{Menara_TCNS2020}).
Let $\Pi = \{\calC_1,\calC_2\}$ be a nontrivial partition of $\calV$.
For $i \in \calC_1$ and $j \in \calC_2$, define
\begin{align*}
  \bar{\omega} := |\omega_i - \omega_j|, \quad \bar{a} := \sum_{l \in \calC_2} a_{il} + \sum_{l \in \calC_1} a_{jl}.
\end{align*}
We assume that $\bar{\omega} > \bar{a}$.
In that case, the solutions to \eqref{eq:linearized_z} is $\bar{T}$-periodic, where $\bar{T} := 2 \pi \bar{\omega}/\sqrt{\bar{\omega}^2 - \bar{a}^2}$ (see Lemma~3.4 in \cite{Menara_TCNS2020}).
The case where $\bar{\omega} < \bar{a}$ corresponds to the well-studied (full) frequency synchronization.

Without loss of generality, we assume that $R_4 = \bbone_{|\calE_\mathrm{inter}|}$.
Then, the Jacobian in \eqref{eq:def_J} can be written by
\begin{align}
  J(\tau,\varepsilon,z_0) := J_\mathrm{intra} + J_\mathrm{inter} \cos(\zeta(\tau,0,z_0,\varepsilon)) \label{eq:Jacobian_two-cluster}
\end{align}
with $J_\mathrm{intra} := \Gamma_1 R_1$ and $J_\mathrm{inter} := \Gamma_2 R_3$.
We note that the Jacobian cannot be written in the above form when the number of clusters is more than two.

The following result provides a structural rather than quantitative condition for two-cluster synchronization.
This condition includes the one in Theorem~3.6 of \cite{Menara_TCNS2020}, where it is assumed that $J_\mathrm{intra} = -cI$ for some $c > 0$.

\begin{theorem} \label{thm:two-cluster}
  Consider the Kuramoto model \eqref{eq:Kuramoto_model_vector} and a nontrivial graph partition $\Pi = \{\calC_1,\calC_2\}$ such that the subgraphs $\calG_1$ and $\calG_2$ are connected.
  Suppose that Assumptions~\ref{ass:natural_frequency} and \ref{ass:external_equitable_partition} hold.
  Suppose also that $\bar{\omega} > \bar{a}$.
  If $J_\mathrm{intra}$ and $J_\mathrm{inter}$ commute, i.e., if $J_\mathrm{intra} J_\mathrm{inter} = J_\mathrm{inter} J_\mathrm{intra}$, then $\calS_\Pi$ is exponentially stable.
\end{theorem}

\begin{IEEEproof}
  The equation \eqref{eq:linearized_z} can now be written by
  \begin{align*}
    \frac{\diff z}{\diff \tau} = 1 - \varepsilon \bar{a} \sin(z).
  \end{align*}
  Since $\varepsilon \bar{a} < 1$, any solution is $\bar{T}$-periodic and has zero average.
  Thus, the average Jacobian is given by $J_\mathrm{intra}$.
  We note that for all $z_0 \in \bbR^m$ and all $\varepsilon > 0$,
  \begin{align}
    &\frac{1}{\bar{T}} \int_\tau^{\tau + \bar{T}} [J(\sigma,\varepsilon,z_0) - J_\mathrm{intra}] \,\diff \sigma \nonumber \\
    &\quad{} = \frac{1}{\bar{T}} \int_\tau^{\tau + \bar{T}} J_\mathrm{inter} \cos(\zeta(\sigma,0,z_0,\varepsilon)) \,\diff \sigma = 0 \label{eq:zero-average}
  \end{align}
  holds uniformly in $\tau$.
  Because $J_\mathrm{intra}$ and $J_\mathrm{inter}$ commute, the equation \eqref{eq:linearization} satisfies the so-called Lappo-Danilevskii condition \cite{Adrianova_book1995}:
  \begin{align*}
    J(\tau,\varepsilon,z_0) \int_{\tau^\prime}^\tau J(\sigma,\varepsilon,z_0) \,\diff \sigma = \int_{\tau^\prime}^\tau J(\sigma,\varepsilon,z_0) \,\diff \sigma J(\tau,\varepsilon,z_0)
  \end{align*}
  for all $\tau,\tau^\prime \in \bbR_+$.
  Thus, the asymptotic stability follows from Theorem~4.2.4 in \cite{Adrianova_book1995}.
  Here, we show that the convergence is uniform with respect to $z_0$.
  Define $\tau_k := k \bar{T}$ for all $k \in \bbZ_+$.
  Let $\Phi(\tau_{k+1},\tau_k)$ denote the state-transition matrix of \eqref{eq:linearization} on $[\tau_k,\tau_{k+1}]$.
  Because the state-transition matrix in the current case can be written as
  \begin{align*}
    &\Phi(\tau_{k+1},\tau_k) \\
    &\quad{} = \e^{\varepsilon \bar{T} J_\mathrm{intra}} + \sum_{l=2}^\infty \frac{\varepsilon^l}{l!} \left( \int_\tau^{\tau + \bar{T}} [J(\sigma,\varepsilon,z_0) - J_\mathrm{intra}] \,\diff \sigma \right)^l,
  \end{align*}
  it follows from \eqref{eq:zero-average} that $\|\Phi(\tau_{k+1},\tau_k)\| < 1$ for all $k \in \bbZ_+$.
  Clearly, this inequality holds uniformly in $z_0$.
  Therefore, we complete the proof by applying Proposition~\ref{prop:nonmonotonic}.
\end{IEEEproof}

\begin{remark}
  Two-cluster synchronization can be considered as a nonlinear version of \emph{bipartite consensus}, which has been studied in the presence of antagonistic interactions \cite{Altafini_TAC2013,Proskurnikov_TAC2016}.
  In this case the underlying network is represented by a signed graph.
  The Kuramoto model on signed graphs was studied in \cite{Delabays_SIADS2019}.
  In this paper we do not consider signed graphs explicitly, but our results can be extended to those cases by using a signed external equitable partition \cite{Schaub_Chaos2016} in place of Assumption~\ref{ass:external_equitable_partition}.
\end{remark}

We present an example where the stability depends on the choice of parameter values.

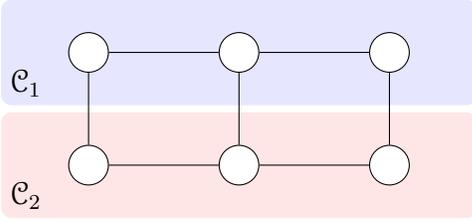
\begin{figure}
  \centering
  \begin{tikzpicture}
    \node [rectangle, fill=blue, fill opacity=0.1, rounded corners, minimum width=180, minimum height=40] (C1) at (0,   0) {};
    \node [rectangle, fill=red,  fill opacity=0.1, rounded corners, minimum width=180, minimum height=40] (C2) at (0,-1.5) {};
    \node [circle, draw=black, fill=white, minimum size=15] (v1) at (-2,   0) {};
    \node [circle, draw=black, fill=white, minimum size=15] (v2) at ( 0,   0) {};
    \node [circle, draw=black, fill=white, minimum size=15] (v3) at ( 2,   0) {};
    \node [circle, draw=black, fill=white, minimum size=15] (v4) at (-2,-1.5) {};
    \node [circle, draw=black, fill=white, minimum size=15] (v5) at ( 0,-1.5) {};
    \node [circle, draw=black, fill=white, minimum size=15] (v6) at ( 2,-1.5) {};
    \draw (v1) -- (v2);
    \draw (v2) -- (v3);
    \draw (v4) -- (v5);
    \draw (v5) -- (v6);
    \draw (v1) -- (v4);
    \draw (v2) -- (v5);
    \draw (v3) -- (v6);
    \node [anchor=south west] (A) at (C1.south west) {\large $\mathcal{C}_1$};
    \node [anchor=south west] (B) at (C2.south west) {\large $\mathcal{C}_2$};
  \end{tikzpicture}
  \caption{Network partitioned into two clusters}
  \label{fig:example2}
\end{figure}

\begin{figure}
  \centering
  \includegraphics[width=\columnwidth]{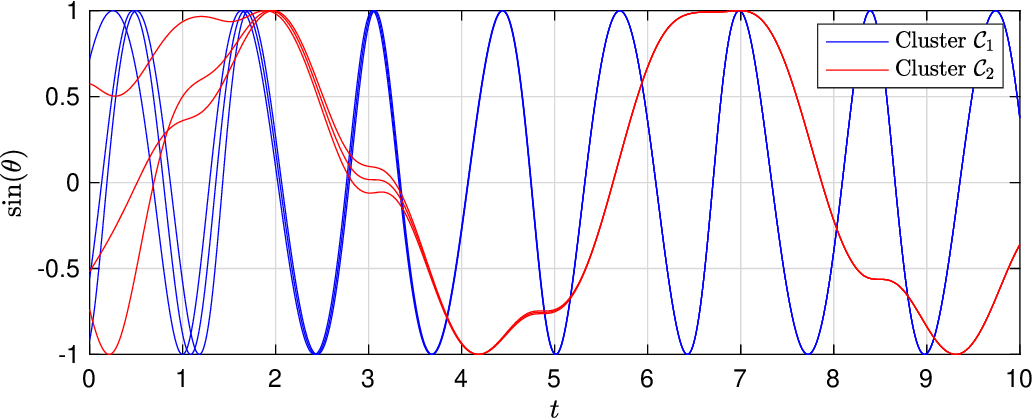} \\
  \vskip 2mm
  \includegraphics[width=\columnwidth]{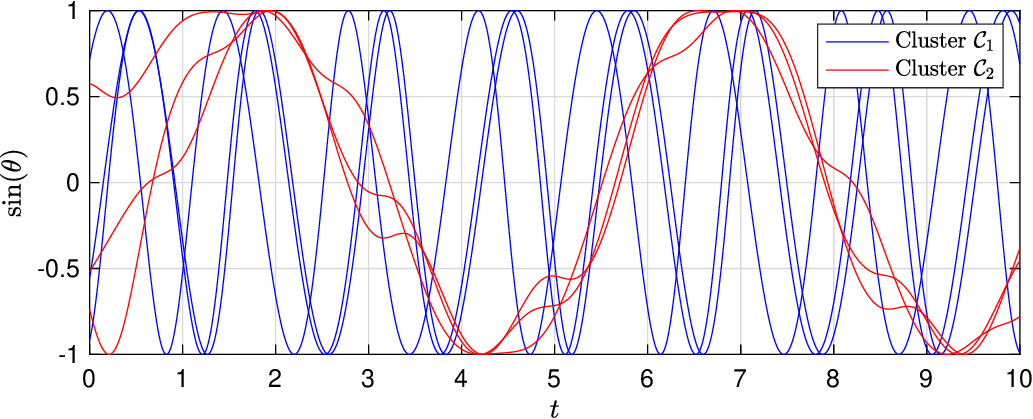}
  \caption{Phase evolutions of the oscillators in Example~\ref{ex:two_clusters} with the stable case (top) and the unstable case (bottom)}
  \label{fig:sim_two_clusters}
\end{figure}

\begin{example} \label{ex:two_clusters}
  Consider the network with two clusters shown in Fig.~\ref{fig:example2}.
  The adjacency matrix is given by
  \begin{align*}
    A = 
    \left[
      \begin{array}{ccc|ccc}
        0 & a_1 & 0 & 0 & 0 & b \\
        a_1 & 0 & a_1 & 0 & b & 0 \\
        0 & a_1 & 0 & b & 0 & 0 \\ \hline
        0 & 0 & b & 0 & a_2 & 0 \\
        0 & b & 0 & a_2 & 0 & a_2 \\
        b & 0 & 0 & 0 & a_2 & 0
      \end{array}
    \right].
  \end{align*}
  We have the Jacobian in \eqref{eq:Jacobian_two-cluster} with
  \begin{align*}
    J_\mathrm{intra} =
    \begin{bmatrix}
      -2a_1 & a_1 & 0 & 0 \\
      a_1 & -2a_1 & 0 & 0 \\
      0 & 0 & -2a_2 & a_2 \\
      0 & 0 & a_2 & -2a_2
    \end{bmatrix},
  \end{align*}
  which is a Hurwitz matrix, and
  \begin{align*}
    J_\mathrm{inter} =
    \begin{bmatrix}
      -b & 0 & 0 & -b \\
      0 & -b & -b & 0 \\
      0 & -b & -b & 0 \\
      -b & 0 & 0 & -b
    \end{bmatrix}.
  \end{align*}
  One can observe that $J_\mathrm{intra}$ and $J_\mathrm{inter}$ commute if $a_1 = a_2$ or $b = 0$.
  Here, we set $a_1 = a_2 = 1$ and $b = 1$.
  The simulation result with $\omega_i = 5$ for $i \in \calC_1$ and $\omega_i = 1$ for $i \in \calC_2$ is shown in Fig.~\ref{fig:sim_two_clusters} (top).
  In this setting, Theorem~\ref{thm:two-cluster} can provide a less conservative stability condition compared with Theorem~\ref{th:multi}, which requires $\bar{\omega} > 125$.
  \textcolor{black}{On the other hand, Theorem~\ref{thm:two-cluster} provides only a sufficient condition for cluster synchronization.}
  Next, we show an unstable case by changing the connectivity in $\calC_1$ to be much smaller by setting $a_1 = 0.01$.
  In this case, $J_\mathrm{intra}$ and $J_\mathrm{inter}$ no longer commute, that is, the condition in Theorem~\ref{thm:two-cluster} does not hold.
  The simulation result with the same natural frequencies is presented in Fig.~\ref{fig:sim_two_clusters} (bottom), which shows that the cluster synchronization manifold is unstable.
\end{example}

\subsection{Cluster Phase Cohesiveness without EEPs} \label{subsec:main-C}

In the previous two subsections, we have considered cluster synchronization under the EEP condition in Assumption~\ref{ass:external_equitable_partition}, which may be practically hard to satisfy.
The objective of this section is to relax this hypothesis, and hence, we do not assume the cluster synchronization manifold to be invariant.

First, we introduce the stability concept to be considered.
Similar definitions of practical stability under perturbations can be found in \cite{Montenbruck_Automatica2015,Lageman_CDC2016}.

\begin{definition}
  The cluster synchronization manifold $\calS_\Pi$ is said to be \emph{$(\alpha,\beta)$-practically stable} for \eqref{eq:Kuramoto_model_vector} if for a given pair $(\alpha,\beta)$ of positive constants, there exists $T(\alpha,\beta) > 0$ such that $\dist(\theta(0),\mathcal{S}_\Pi) < \alpha$ implies $\dist(\theta(t),\mathcal{S}_\Pi) < \beta$ for all $t \ge T(\alpha,\beta)$.
\end{definition}

Because the linear approximation in Section~\ref{subsec:partial_linearization} cannot be applied, we need to directly address the nonlinear system in \eqref{eq:dxdtau} and \eqref{eq:dzdtau}.
In particular, we are concerned with ultimate boundedness of solutions with respect to $x$ uniformly in $z$ (see \cite[Chap.~4]{Haddad-Chellaboina_book2008}).
To do so, we consider the differential equation
\begin{align}
  \frac{\diff x}{\diff \tau} = \varepsilon f(x,\zeta(\tau,x_0,z_0,\varepsilon)), \quad x(0) = x_0. \label{eq:dxdtau2}
\end{align}
Note that the right-hand side depends explicitly on the initial states $x_0$ and $z_0$.
Thus, depending on the initial states, the right-hand side of \eqref{eq:dxdtau2} changes.

Now, we provide some preliminary results.
Recall that $z = \zeta(\tau,x_0,z_0,\varepsilon)$ denotes the solutions to \eqref{eq:dzdtau} with respect to the initial states $x_0$ and $z_0$ in \eqref{eq:dxdtau} and \eqref{eq:dzdtau}, respectively.
If $\varepsilon = 0$, we can explicitly solve the equation \eqref{eq:dzdtau}.
Then, we can define the average vector field
\begin{align*}
  f_\mathrm{av}(x) &:= \lim_{T \to \infty} \frac{1}{T} \int_\tau^{\tau + T} f(x,\zeta(\sigma,x_0,z_0,0)) \,\diff \sigma \\
  &= \Gamma_1 \sin(R_1 x).
\end{align*}
This is well defined uniformly in $\tau$ as well as in $x_0$ and $z_0$.
Let $L > 0$ be a Lipschitz constant of $f_\mathrm{av}$.
The following lemma characterizes the closeness between the solutions of \eqref{eq:dxdtau2} and those of the average system.

\begin{lemma} \label{lem:integral-bound2}
  There exists a constant $K > 0$ such that $K \to 0$ as $\|W_\mathrm{inter}\|_\infty \to 0$ and for all $x \in \bbT^n$, all $T > 0$, and all $\varepsilon > 0$,
  \begin{align*}
    \left\| \frac{1}{T} \int_\tau^{\tau + T} [f(x,\zeta(\sigma,x_0,z_0,\varepsilon)) - f_\mathrm{av}(x)] \,\diff \sigma \right\| \le K \left( \frac{1}{T} + \varepsilon \right)
  \end{align*}
  holds uniformly in $\tau$, $x_0$, and $z_0$.
\end{lemma}

\begin{IEEEproof}
  See Appendix~\ref{app:integral-bound2}.
\end{IEEEproof}

We are now ready to present our main results of this subsection.

\begin{theorem} \label{th:main3}
  Consider the Kuramoto model \eqref{eq:Kuramoto_model_vector} and a nontrivial graph partition $\Pi = \{\calC_1,\ldots,\calC_r\}$ such that the subgraphs $\calG_1,\ldots,\calG_r$ are all connected.
  Suppose that Assumption~\ref{ass:natural_frequency} holds.
  Then, the following statements are true:
  \begin{enumerate}
    \item For each $\beta > 0$, there exists $a^* > 0$ such that if
    \begin{align*}
      a_{ij} < a^*, \quad (i,j) \in \calE_\mathrm{inter},
    \end{align*}
    then $\calS_\Pi$ is $(\alpha,\beta)$-practically stable with some $\alpha > 0$.
    \item For each $\beta > 0$, there exists $\omega^* > 0$ such that if
    \begin{align*}
      |\omega_i - \omega_j| > \omega^*, \quad (i,j) \in \calE_\mathrm{inter},
    \end{align*}
    then $\calS_\Pi$ is $(\alpha,\beta)$-practically stable with some $\alpha > 0$.
  \end{enumerate}
\end{theorem}

\begin{IEEEproof}
  According to Theorem~\ref{thm:nonlinear_averaging}, we know that corresponding to any $\beta > 0$, there exists $\varepsilon^* > 0$ such that if $\varepsilon < \varepsilon^*$, then for an appropriate choice of $T(\beta) > 0$, $\|x(t)\| < \beta$ for all $t \ge T(\beta)$ whenever the initial state $x(0)$ lies in a small neighborhood of the origin.
  Thus, the second part of the theorem is a direct consequence of Theorem~\ref{thm:nonlinear_averaging}.
  To prove the first part, we can utilize Lemma~\ref{lem:integral-bound2}.
  Specifically, $\Omega$ in the proof of Theorem~\ref{thm:nonlinear_averaging} approaches zero as $K \to 0$.
  This implies the desired result.
\end{IEEEproof}

Theorem~\ref{th:main3} indicates that, under the same conditions as those in Theorem~\ref{th:multi}, clustered patterns of synchronization can arise even if Assumption~\ref{ass:external_equitable_partition} is not satisfied.
In particular, if the inter-cluster coupling strengths are weak or if the inter-cluster natural frequency differences are large, then the synchronization pattern becomes apparent.
\textcolor{black}{A closely related result can be found in \cite{Menara_CDC2019,Qin_TAC2021}; however, the effects of the natural frequency differences between clusters were not addressed.
The main contribution of Theorem~\ref{th:main3} is to have clarified how the heterogeneous natural frequencies affect the phase cohesiveness within clusters.}

\section{Case Study with Brain Networks} \label{sec:brain}

This section focuses on applications of the results derived so far to brain networks.
\textcolor{black}{The presentations here are mostly motivated by \cite{Cabral2011,Menara_CDC2019}.}

\subsection{Basic Setting}

Mapping of human brain networks with noninvasive methods has been an active research area in the neuroscience community \cite{Sporns_book2011}.
We consider the brain networks identified in \cite{Hagmann_plosb2008} for 5 human subjects, whose data including their weighted adjacency matrices can be found at the USC Multimodal Connectivity Database \cite{Brown12}.
Those networks were extracted with diffusion spectrum imaging and consist of 998 cortical regions of equal sizes, which are further classified into anatomically determined 66 regions.

\begin{figure}
  \centering
  \vspace{-4mm}
  \includegraphics[width=\columnwidth]{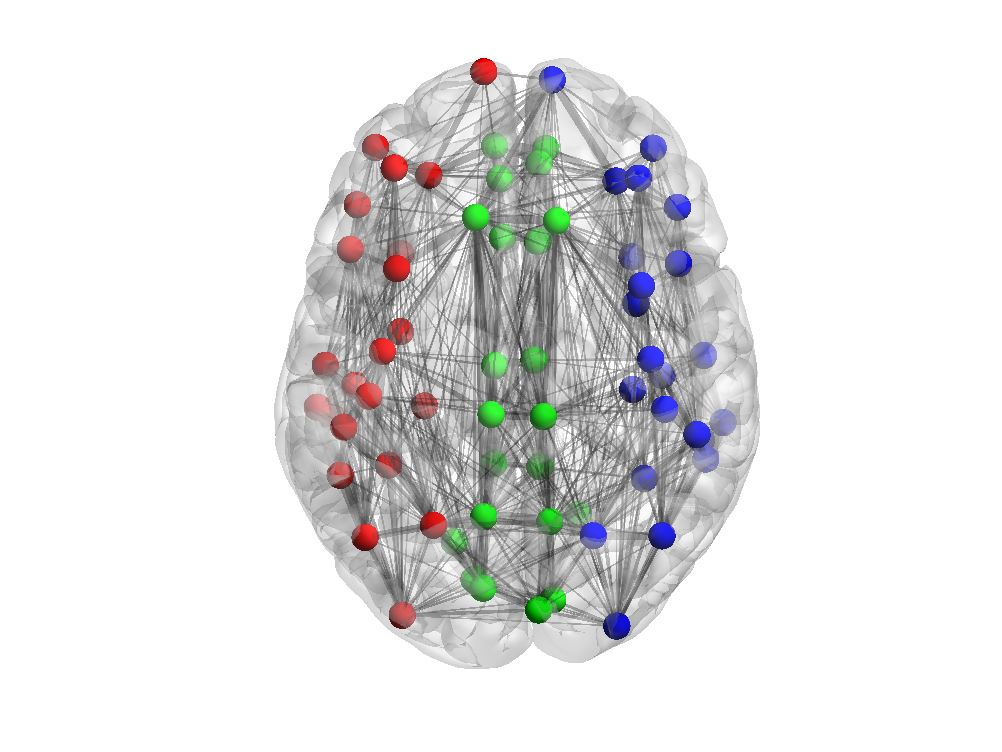}
  \vspace{-10mm}
  \caption{Network structure of the brain}
  \label{fig:brain}
\end{figure}

\begin{figure}
  \centering
  \includegraphics[width=0.8\columnwidth]{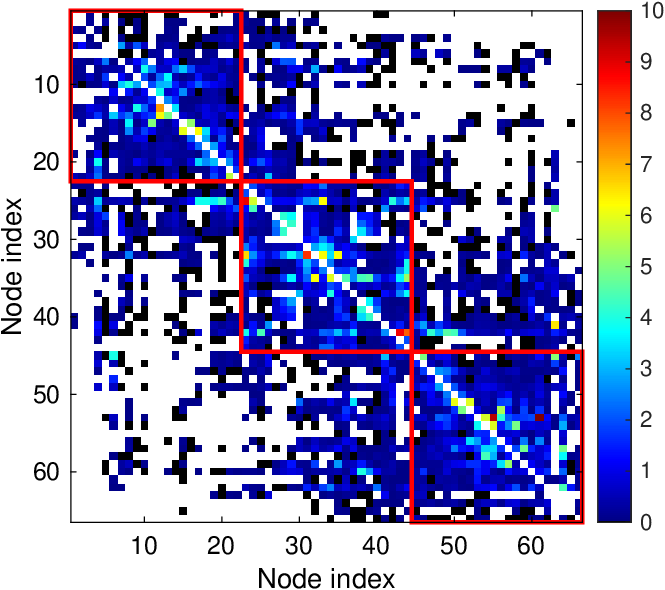}
  \caption{Adjacency matrix of the structural brain network}
  \label{fig:adjacency}
\end{figure}

Following \cite{Honey_PNAS2009,Cabral2011}, we process the empirical data to construct a representative network for computer simulation.
We first binarize all the 5 weighted adjacency matrices and then average them into the 66 regions, that is, sum up all the weights in each region and divide it by the number of the target regions.
After that, we average the 66-region networks of 5 human subjects and normalize it so that the weights belong to the range $[0,10]$.
The obtained representative network is shown in Fig.~\ref{fig:brain}, where the image was generated by BrainNet Viewer \cite{Xia_ploso2013} and only 30\% of edges are visualized.
We also present in Fig.~\ref{fig:adjacency} the obtained adjacency matrix.
Note that this network is nonsymmetric but is close to a symmetric one.
In the following, we consider three clusters as presented in Figs.~\ref{fig:brain} and \ref{fig:adjacency}, where each cluster consists of 22 nodes.

\textcolor{black}{There are many types of mathematical models representing neural activities in the literature.
At the macroscopic level, the so-called Wilson--Cowan model consists of the populations of neurons with excitatory and inhibitory types.
This model represents oscillatory phenomena in the phase space of the mean numbers of activated excitatory and inhibitory neurons \cite{Amari_book,Wilson_Biophys1972}.
The Kuramoto model can approximate the Wilson--Cowan model \cite{Sadilek_2015}.
Hence, as also done in \cite{Cabral2011}, we simply use the Kuramoto model in the subsequent simulations.}

\subsection{Local and Global Order Parameters}

\textcolor{black}{In addition to structural connectivity discussed above, functional connectivity in brain networks is important to understand the mechanisms of human brains (see \cite{Sporns_book2011} and the references therein).
Functional connectivity is measured as the correlations of neural activities among the network and exhibits characteristic patterns depending on the brain state.
Following \cite{Cabral2011}, we suppose that the brain network produces synchronization patterns with the three clusters as indicated in Figs.~\ref{fig:brain} and \ref{fig:adjacency}.
Here, we demonstrate that our theoretical findings on cluster synchronization can also characterize functional connectivity in brain networks.}

To measure the synchronization level in each of the three clusters, we define the cluster-wise order parameters as
\begin{align*}
  r_p(t) = \left| \frac{1}{\mathrm{card}(\calC_p)} \sum_{i \in \calC_p} \e^{j \theta_i(t)} \right|, \quad p \in \{1,2,3\}.
\end{align*}
In geometric interpretations, the order parameter represents the distance from the origin to the mean position of the oscillators on the unit circle.
That is, $r_p(t) = 1$ when all phases of the oscillators in the cluster $\calC_p$ are positioned at the same point, and $r_p(t) = 0$ when they are well spread with equal distances.
In addition, to quantify the synchronization level among the whole network, we define the network-wide order parameter
\begin{align*}
  r(t) = \left| \frac{1}{N} \sum_{i=1}^N \e^{j \theta_i(t)} \right|.
\end{align*}

\begin{figure}
  \centering
  \includegraphics[width=0.9\columnwidth]{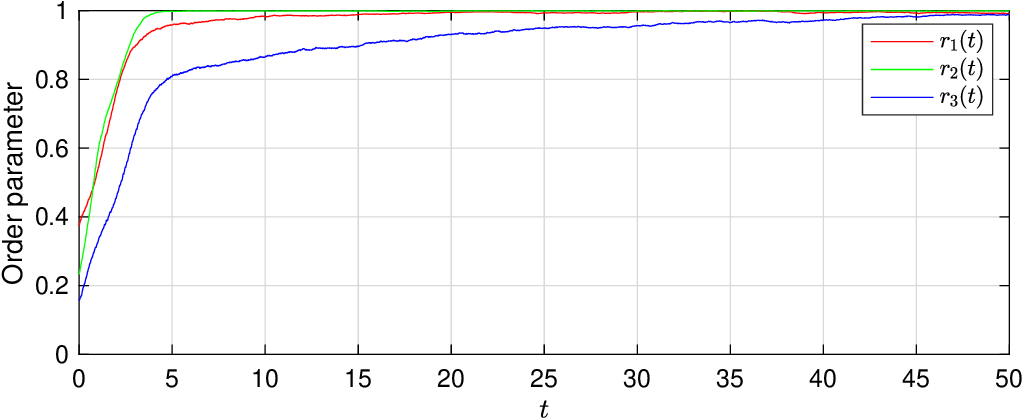} \\
  \vskip 2mm
  \includegraphics[width=0.9\columnwidth]{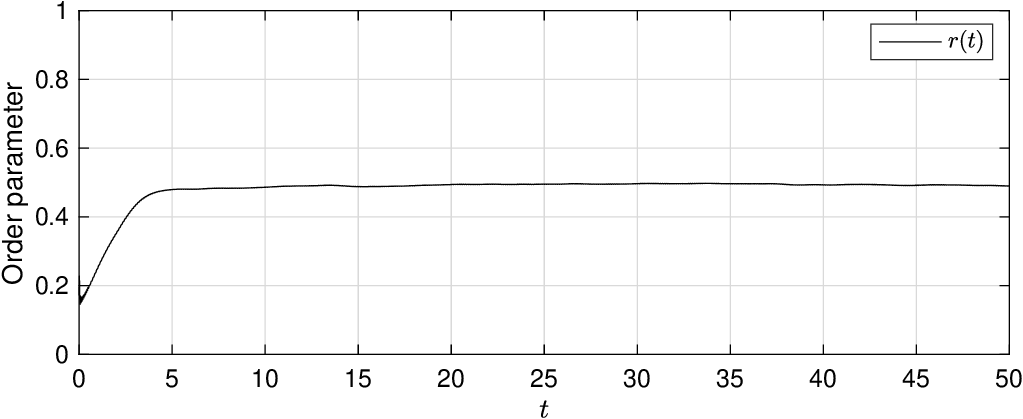}
  \caption{Simulation results for the cluster-wise order parameters (top) and the network-wide order parameter (bottom)}
  \label{fig:order_paras}
\end{figure}

In the Kuramoto model, we set the natural frequencies in the Kuramoto model as follows.
\textcolor{black}{In \cite{Cabral2011}, the authors produce the oscillators' frequencies by the Gaussian distribution around 60 Hz.}
To show the effects of the oscillators' heterogeneity, we set the frequencies as $50$, $60$, and $70$ Hz for the clusters $\calC_1$, $\calC_2$, and $\calC_3$, respectively.
Remark that we do not consider any delay effect in this paper through it may also be a key factor in functional connectivity.
Also, we add the Gaussian noise with standard derivation $2$ rad/s in the Kuramoto model.
The simulation result is presented in Fig.~\ref{fig:order_paras}, where the order parameters introduced above are shown.
The data are smoothed over 10 seconds for the network-wide order parameter.
As shown in those figures, the local order parameters in clusters are moderately high while the global order parameter is low.
Therefore, cluster synchronization patterns can be observed.

\begin{figure}
  \centering
  \includegraphics[width=0.9\columnwidth]{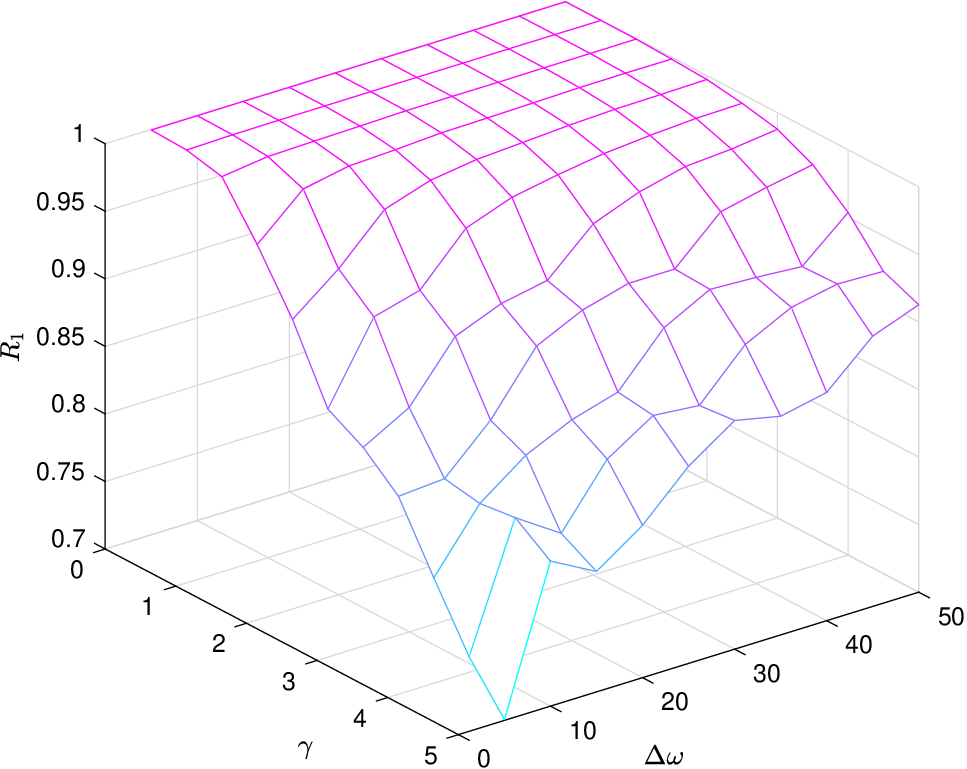}
  \caption{Relations among the inter-cluster coupling strengths, the inter-cluster natural frequency differences, and the cluster-wise order parameter}
  \label{fig:order_paras_tradeoffs}
\end{figure}

\textcolor{black}{Next, we aim to show the interrelation among the network parameters and functional connectivity measured by the order parameters.
We present in Fig.~\ref{fig:order_paras_tradeoffs} the order parameter in the cluster $\calC_1$ averaged over time, denoted $R_1$, for several coupling strengths and natural frequencies.
In particular, the inter-cluster edge weights in the adjacency matrix in Fig.~\ref{fig:adjacency} are multiplied by the scaling factor $\gamma$ and the inter-cluster natural frequency differences are set by $\Delta \omega$.
This demonstrates the theoretical results in Section~\ref{sec:main}.
That is, weak connections and heterogeneous intrinsic dynamics among clusters help the functional connectivity to be strong.}

Finally, we give some discussions.
In the neuroscience community, functional connectivity has been studied with graph-theoretic analysis \cite{Bullmore_NatRevNeurosci2009} and numerical analysis \cite{Deco_PNAS2009}.
On the other hand, this paper has provided a theoretical framework based on oscillator models to explain the appearance of robust patterns in functional connectivity.
Specifically, as demonstrated above, the structural network connectivity and the intrinsic dynamics contain significant information for cluster synchronization.
It would be of interest to analyze functional brain networks from these viewpoints and moreover see if healthy and pathological states of brains may be characterized based on this criterion.

\section{Conclusions and Future Directions} \label{sec:conclusion}

We have shown a unified framework for stability analysis to study cluster synchronization and cluster phase cohesiveness of Kuramoto oscillators.
Our conditions are different from those derived in the previous works in how the heterogeneity of natural frequencies affects the (strict) stability properties.
Also, we have established the extended method of averaging in stability theory.
Our findings can provide theoretical insights in functional connectivity of brain networks, which has recently attracted much interests in the neuroscience community.

Some directions for future research can be considered.
On one hand, practical control methodologies for cluster synchronization should be developed (see also \cite{Menara_CDC2019} for the motivation from neuroscience).
On the other hand, interactions with delays may induce remote synchronization, where oscillators that are not directly connected synchronize \cite{Punetha_2015}.
However, we have only limited knowledge about its mechanisms especially related to the averaging principle (see \cite{Qin_TAC2021b} for a simple case).

\appendix

\subsection{Derivation of $R_1$, $R_2$, $R_3$, and $R_4$} \label{app:derivation}

Since $\tilde{B}$ has full column rank, we have $R = B^\trans (\tilde{B}^\trans)^\dagger = B^\trans (\tilde{B}^\dagger)^\trans$.
The problem is thus to calculate the Moore--Penrose inverse $\tilde{B}^\dagger$ of the column-wise partitioned matrix $\tilde{B}$.
To this end, we borrow a useful lemma from \cite{Baksalary2007}.

\begin{lemma} \label{lem:psendoinverse}
  Let $M_1 \in \bbR^{m \times n_1}$ and $M_2 \in \bbR^{m \times n_2}$.
  Define the orthogonal projectors $P_1,P_2 \in \bbR^{m \times m}$ by $P_i := I_m - M_i M_i^\dagger$ for $i \in \{1,2\}$.
  Then, we have
  \begin{align*}
    \begin{bmatrix}
      M_1 & M_2
    \end{bmatrix}^\dagger =
    \begin{bmatrix}
      (P_2 M_1)^\dagger \\
      (P_1 M_2)^\dagger
    \end{bmatrix}
  \end{align*}
  if and only if $\image M_1 \cap \image M_2 = \{0\}$.
\end{lemma}

Let $P := I_N - \tilde{B}_\mathrm{intra} \tilde{B}_\mathrm{intra}^\dagger$ and $Q := I_N - \tilde{B}_\mathrm{inter} \tilde{B}_\mathrm{inter}^\dagger$.
Note that these are orthogonal projectors onto $\kernel \tilde{B}_\mathrm{intra}^\trans$ and $\kernel \tilde{B}_\mathrm{inter}^\trans$, respectively.
Since $\image \tilde{B}_\mathrm{intra} \cap \image \tilde{B}_\mathrm{inter} = \{0\}$, Lemma~\ref{lem:psendoinverse} provides
\begin{align*}
  \tilde{B}^\dagger =
  \begin{bmatrix}
    \tilde{B}_\mathrm{intra} & \tilde{B}_\mathrm{inter}
  \end{bmatrix}^\dagger
  =
  \begin{bmatrix}
    (Q \tilde{B}_\mathrm{intra})^\dagger \\
    (P \tilde{B}_\mathrm{inter})^\dagger
  \end{bmatrix}.
\end{align*}
It thus follows that
\begin{align*}
  R &=
  \begin{bmatrix}
    B_\mathrm{intra}^\trans \\
    B_\mathrm{inter}^\trans
  \end{bmatrix}
  \begin{bmatrix}
    (\tilde{B}_\mathrm{intra}^\trans Q)^\dagger & (\tilde{B}_\mathrm{inter}^\trans P)^\dagger
  \end{bmatrix} \\
  &=
  \begin{bmatrix}
    B_\mathrm{intra}^\trans (\tilde{B}_\mathrm{intra}^\trans Q)^\dagger & B_\mathrm{intra}^\trans (\tilde{B}_\mathrm{inter}^\trans P)^\dagger \\
    B_\mathrm{inter}^\trans (\tilde{B}_\mathrm{intra}^\trans Q)^\dagger & B_\mathrm{inter}^\trans (\tilde{B}_\mathrm{inter}^\trans P)^\dagger
  \end{bmatrix}.
\end{align*}
We notice that $\kernel B_\mathrm{intra}^\trans = \kernel \tilde{B}_\mathrm{intra}^\trans$, which follows from the fact that $B_\mathrm{intra}^\trans \bbone_n = 0$ and $\tilde{B}_\mathrm{intra}^\trans \bbone_n = 0$.
Hence, we have $B_\mathrm{intra}^\trans (\tilde{B}_\mathrm{inter}^\trans P)^\dagger = 0$ since $\image P = \kernel B_\mathrm{intra}^\trans$.

\subsection{Proof of Theorem~\ref{thm:linear_averaging}} \label{app:linear_averaging}

The proof mostly follows from \cite{Stilwell_SIADS2006}.
Since $A_\mathrm{av}$ is a Hurwitz matrix, there exists a positive-definite matrix $P = P^\trans$ such that $A_\mathrm{av}^\trans P + PA_\mathrm{av}\prec 0$.
Thus, the positive-definite function $V(x) = x^\trans Px$ possesses the following properties:
\begin{enumerate}
  \item there exist $c_1,c_2 > 0$ such that for all $x \in \bbR^n$,
  \begin{align*}
    c_1 \|x\|^2 \le V(x) \le c_2 \|x\|^2;
  \end{align*}
  \item there exists $c_3 > 0$ such that for all $x \in \bbR^n$,
  \begin{align*}
    \frac{\partial V}{\partial x}(x) A_\mathrm{av} x \le -c_3 \|x\|^2.
  \end{align*}
\end{enumerate}

Given $t_0 \in \bbR_+$, we define $t_k := t_0 + kT$ for $k \in \bbZ_+$, where $T > 0$ is determined later.
Let $\Phi(t_{k+1},t_k)$ denote the state-transition matrix of \eqref{eq:linear-averaging} over $[t_k,t_{k+1}]$, which can be expanded to the Peano--Baker series \cite{Rugh_book1996}:
\begin{align*}
  \Phi(t_{k+1},t_k) = I + \varepsilon \int_{t_k}^{t_{k+1}} A(s,\varepsilon) \,\diff s + R_k(\varepsilon),
\end{align*}
where
\begin{align*}
  R_k(\varepsilon) := \sum_{l=2}^\infty \varepsilon^l \int_{t_k}^{t_{k+1}} A(s_1,\varepsilon) \cdots \int_{t_k}^{s_{l-1}} A(s_l,\varepsilon) \,\diff s_l \cdots \diff s_1.
\end{align*}
By defining $\alpha := \sup \{\|A(t,\varepsilon)\| : t \ge 0,\ \varepsilon \in (0,\varepsilon_0]\}$, which is finite since $A$ is assumed to be bounded, we have
\begin{align}
  &\|R_k(\varepsilon)\| \nonumber \\
  &\quad{} \le \sum_{l=2}^\infty \varepsilon^l \int_{t_k}^{t_{k+1}} \|A(s_1,\varepsilon)\| \cdots \int_{t_k}^{s_{l-1}} \|A(s_l,\varepsilon)\| \,\diff s_l \cdots \diff s_1 \nonumber \\
  &\quad{} \le \e^{\alpha \varepsilon T} - 1 - \alpha \varepsilon T. \label{eq:R_bound}
\end{align}
Let $\Phi_\mathrm{av}(t_{k+1},t_k)$ be the state-transition matrix of the average system over $[t_k,t_{k+1}]$, which can be expanded as
\begin{align*}
  \Phi_\mathrm{av}(t_{k+1},t_k) = I + \varepsilon T A_\mathrm{av} + \sum_{l = 2}^\infty \frac{(\varepsilon T A_\mathrm{av})^l}{l!}.
\end{align*}
Note that the norm of the summation in the right-hand side has the same bound as in \eqref{eq:R_bound}.
Now, we define $H(t_{k+1},t_k) := \Phi(t_{k+1},t_k) - \Phi_\mathrm{av}(t_{k+1},t_k)$.
It then follows from \eqref{eq:integral} that
\begin{align*}
  \|H(t_{k+1},t_k)\| \le \varepsilon T [\sigma(T) + \rho(\varepsilon)] + 2 (\e^{\alpha \varepsilon T} - 1 - \alpha \varepsilon T).
\end{align*}

Again, we consider the function $V(x)$.
Along the solutions to \eqref{eq:linear-averaging}, we have
\begin{align*}
  &V(x(t_{k+1})) - V(x(t_k)) \\
  &\quad{} = x(t_k)^\trans [\Phi(t_{k+1},t_k)^\trans P \Phi(t_{k+1},t_k) - P] x(t_k) \\
  &\quad{} = x(t_k)^\trans [\Phi_\mathrm{av}(t_{k+1},t_k)^\trans P \Phi_\mathrm{av}(t_{k+1},t_k) - P \\
  &\qquad{} + 2 \Phi_\mathrm{av}(t_{k+1},t_k)^\trans P H(t_{k+1},t_k) \\
  &\qquad{} + H(t_{k+1},t_k)^\trans P H(t_{k+1},t_k)] x(t_k).
\end{align*}
For the first two terms in the far right-hand side, it can be observed that
\begin{align*}
  &x(t_k)^\trans [\Phi_\mathrm{av}(t_{k+1},t_k)^\trans P \Phi_\mathrm{av}(t_{k+1},t_k) - P] x(t_k) \\
  &\quad{} \le c_2 (\e^{- \frac{c_3}{c_2} \varepsilon T} - 1) \|x(t_k)\|^2.
\end{align*}
Note that $\|P\| \le c_2$ and $\|\Phi_\mathrm{av}(t_{k+1},t_k)\| \le \sqrt{c_2/c_1} \e^{- \frac{c_3}{2 c_2} \varepsilon T}$.
Thus, we have
\begin{align}
  &V(x(t_{k+1})) - V(x(t_k)) \nonumber \\
  &\le c_2 \left[ (\e^{- \frac{c_3}{c_2} \varepsilon T} - 1) + 2 \sqrt{\frac{c_2}{c_1}} \e^{- \frac{c_3}{2 c_2} \varepsilon T} h(\varepsilon,T) + h(\varepsilon,T)^2 \right] \nonumber \\
  &\quad{} \times \|x(t_k)\|^2, \label{eq:difference-of-V}
\end{align}
where $h(\varepsilon,T) := \varepsilon T [\sigma(T) + \rho(\varepsilon)] + 2 (\e^{\alpha \varepsilon T} - 1 - \alpha \varepsilon T)$.
We define $\tilde{h}(\varepsilon,T)$ by the coefficient of $\|x(t_k)\|^2$ in the right-hand side of \eqref{eq:difference-of-V},
As $\tilde{h}(0,T) = 0$, it can be verified that
\begin{align*}
  \limsup_{\varepsilon \to 0^+} \frac{\tilde{h}(\varepsilon,T) - \tilde{h}(0,T)}{\varepsilon} = - T \left( c_3 - 2 c_2 \sqrt{\frac{c_2}{c_1}} \sigma(T) \right).
\end{align*}
Because $\sigma$ is of class $\mathcal{L}$, we can set $T$ for which $c_3 - 2 c_2 \sqrt{c_2/c_1} \sigma(T) > 0$.
This implies that for sufficiently small $\varepsilon > 0$, we have $\tilde{h}(\varepsilon,T) < 0$.
As a result, there exists $\gamma > 0$ such that
\begin{align*}
  \frac{1}{T} [V(x(t_{k+1})) - V(x(t_k))] \le - \gamma \|x(t_k)\|^2
\end{align*}
for all $k \in \bbZ_+$.
It is clear that $\gamma$ can be chosen independent of $t_0$.
Therefore, by choosing $\varepsilon$ and $T$ as above, $V$ satisfies the hypotheses in Proposition~\ref{prop:nonmonotonic}.
The proof is complete. \hfill \IEEEQED

\subsection{Proof of Theorem~\ref{thm:nonlinear_averaging}} \label{app:nonlinear_averaging}

First, we prove the following lemma.

\begin{lemma} \label{lem:auxiliary}
  For a fixed $t_0 \in \bbR_+$, let $x \colon [t_0,\infty) \to \bbR^n$ and $y \colon [t_0,\infty) \to \bbR^n$ respectively be solutions to \eqref{eq:nonlinear} and \eqref{eq:nonlinear-average} starting in $D$.
  If $x(t_0) = y(t_0)$, then
  \begin{align*}
    \|x(t) - y(t)\| \le K \varepsilon [1 + \varepsilon (t - t_0)] \e^{L \varepsilon (t - t_0)}
  \end{align*}
  for all $t \in [t_0,t_f)$, where $t_f$ is time at which $x(t)$ or $y(t)$ reaches the boundary of $D$.
  If $x(t)$ and $y(t)$ remain in $D$, then $t_f = \infty$.
\end{lemma}

\begin{IEEEproof}
  The solution $x$ to \eqref{eq:nonlinear} follows
  \begin{align*}
    \dot{x}(t) = \varepsilon f_\mathrm{av}(x(t)) + \varepsilon [f(t,x(t),\varepsilon) - f_\mathrm{av}(x(t))].
  \end{align*}
  By integrating the both sides over $[t_0,t]$, we obtain
  \begin{align*}
    x(t) &= x(t_0) + \varepsilon \int_{t_0}^t f_\mathrm{av}(x(s)) \,\diff s \\
    &\quad{} + \varepsilon \int_{t_0}^t [f(s,x(s),\varepsilon) - f_\mathrm{av}(x(s))] \,\diff s.
  \end{align*}
  Also, by integrating \eqref{eq:nonlinear-average}, we have
  \begin{align*}
    y(t) = y(t_0) + \varepsilon \int_{t_0}^t f_\mathrm{av}(y(s)) \,\diff s.
  \end{align*}
  Since $x(t_0) = y(t_0)$, it follows that
  \begin{align*}
    \|x(t) - y(t)\| &\le \varepsilon \int_{t_0}^t \|f_\mathrm{av}(x(s)) - f_\mathrm{av}(y(s))\| \,\diff s \\
    &\quad{} + \varepsilon \left\| \int_{t_0}^t [f(s,x(s),\varepsilon) - f_\mathrm{av}(x(s))] \,\diff s \right\|.
  \end{align*}
  By the Lipschitz continuity and the hypothesis, we obtain
  \begin{align*}
    \|x(t) - y(t)\| &\le L \varepsilon \int_{t_0}^t \|x(s) - y(s)\| \,\diff s \\
    &\quad{} + K \varepsilon [1 + \varepsilon (t - t_0)].
  \end{align*}
  Then, Gr{\"o}nwall's inequality yields
  \begin{align*}
    \|x(t) - y(t)\| \le K \varepsilon [1 + \varepsilon (t - t_0)] \e^{L \varepsilon (t - t_0)}.
  \end{align*}
  This inequality holds as long as $x(t)$ and $y(t)$ remain in $D$.
  The proof is complete.
\end{IEEEproof}

We now prove the theorem.
By the converge theorem, there exist a constant $\delta > 0$ and a $C^1$ function $V \colon \mathcal{B}_\delta \to \bbR_+$ such that the following properties are satisfied:
\begin{enumerate}
  \item there exist $c_1,c_2 > 0$ such that for all $x \in \mathcal{B}_\delta$,
  \begin{align*}
    c_1 \|x\|^2 \le V(x) \le c_2 \|x\|^2;
  \end{align*}
  \item there exists $c_3 > 0$ such that for all $x \in \mathcal{B}_\delta$,
  \begin{align*}
    \frac{\partial V}{\partial x}(x) f_\mathrm{av}(x) \le -c_3 \|x\|^2.
  \end{align*}
\end{enumerate}

Consider the time sequence defined by $t_k := t_0 + kT$ for $k \in \bbZ_+$, where $T > 0$ is determined later.
First, we show that there exists $\delta^\prime \in (0,\delta)$ such that if $x(t_k) \in \mathcal{B}_{\delta^\prime}$, then $x(t_{k+1}) \in \mathcal{B}_\delta$.
Let $x$ and $y$ respectively denote the solutions to \eqref{eq:nonlinear} and \eqref{eq:nonlinear-average} such that $x(t_k) = y(t_k)$.
It follows from Lemma~\ref{lem:auxiliary} that
\begin{align*}
  \|x(t_{k+1}) - y(t_{k+1})\| \le K \varepsilon (1 + \varepsilon T) \e^{L \varepsilon T}.
\end{align*}
We note that $\|y(t_{k+1})\| \le \sqrt{c_2/c_1} \e^{-c_3 \varepsilon T} \|x(t_k)\|$.
Thus,
\begin{align*}
  \|x(t_{k+1})\| &\le \|y(t_{k+1})\| + \|x(t_{k+1}) - y(t_{k+1})\| \\
  &\le \sqrt{\frac{c_2}{c_1}} \e^{-c_3 \varepsilon T} \|x(t_k)\| + K \varepsilon (1 + \varepsilon T) \e^{L \varepsilon T}.
\end{align*}
Let us fix $\varepsilon T = \ell$.
Then, for sufficiently small $\varepsilon$, we can take $\delta^\prime$ mentioned above by solving the inequality
\begin{align*}
  \sqrt{\frac{c_2}{c_1}} \e^{-c_3 \ell} \delta^\prime + K \varepsilon (1 + \ell) \e^{L \ell} < \delta.
\end{align*}


Next, we consider the positive-definite function $V$ introduced above.
We here assume that $x(t_k) \in \mathcal{B}_{\delta^\prime}$ and note that
\begin{align*}
  V(x(t_{k+1})) &= V(y(t_{k+1})) + V(x(t_{k+1})) - V(y(t_{k+1})) \\
  &\le \e^{-c_3 \ell} V(x(t_k)) + \Delta V(x(t_{k+1}),y(t_{k+1})),
\end{align*}
where $\Delta V(x,y) := V(x) - V(y)$.
By the property 1) of $V$, we obtain
\begin{align*}
  \Delta V(x,y) &= V(x - y + y) - V(y) \\
  &\le c_2 \|(x - y) + y\|^2 - V(y) \\
  &\le 2c_2 (\|x - y\|^2 + \|y\|^2) - V(y) \\
  &\le \frac{2c_2}{c_1} V(x - y) + \left( \frac{2c_2}{c_1} - 1 \right) V(y).
\end{align*}
This implies that
\begin{align*}
  &V(x(t_{k+1})) - V(x(t_k)) \\
  &\quad{} \le (\e^{-c_3 \ell} - 1) V(x(t_k)) + \frac{2c_2}{c_1} V(x(t_{k+1}) - y(t_{k+1})) \\
  &\qquad{} + \left( \frac{2c_2}{c_1} - 1 \right) \e^{-c_3 \ell} V(y(t_k)).
\end{align*}
Because $x(t_k) = y(t_k)$ and
\begin{align*}
  V(x(t_{k+1}) - y(t_{k+1})) \le c_2 K^2 \varepsilon^2 (1 + \ell)^2 \e^{2L \ell},
\end{align*}
we obtain
\begin{align*}
  &V(x(t_{k+1})) - V(x(t_k)) \\
  &\quad{} \le \left( \frac{2c_2}{c_1} \e^{-c_3 \ell} - 1 \right) V(x(t_k)) + \frac{2c_2^2}{c_1} K^2 \varepsilon^2 (1 + \ell)^2 \e^{2L \ell}.
\end{align*}
Recall that $\varepsilon T = \ell$ is fixed.
We now choose $T$ such that $(2c_2/c_1) \e^{-c_3 \ell} - 1 < 0$.

From the above discussions, there exist $c,d > 0$ such that
\begin{align*}
  &V(x(t_{k+1})) - V(x(t_k)) \le -c \|x(t_k)\|^2 + d \varepsilon^2
\end{align*}
for all $k \in \bbZ_+$ for which $x(t_k) \in \mathcal{B}_{\delta^\prime}$.
For example, we can take $c = (1/c_2) [1 - (2c_2/c_1) \e^{-c_3 \ell}]$ and $d = (2c_2^2/c_1) K^2 (1 + \ell)^2 \e^{2L \ell}$.
This means that if $\|x(t_k)\|^2 \ge (d/c) \varepsilon^2$, then the value of $V(x(t_{k+1}))$ does not increase from $V(x(t_k))$.
More precisely, we can take $\Omega > \varepsilon \sqrt{d/c}$ so that the strict inequality $\|x(t_k)\|^2 > (d/c) \varepsilon^2$ holds.
We can assume without loss of generality that $\Omega < \delta^\prime$ for sufficiently small $\varepsilon$.
As a result, there exists $\gamma > 0$ such that $\|x(t_k)\| \ge \Omega$ implies
\begin{align}
  V(x(t_{k+1})) - V(x(t_k)) \le - \gamma \|x(t_k)\|^2. \label{eq:Delta_V}
\end{align}

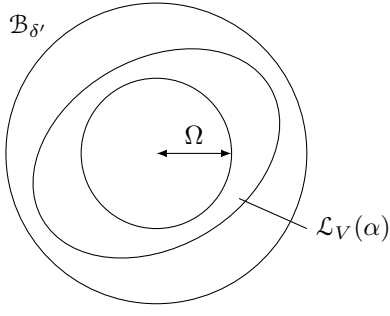
\begin{figure}
  \centering
  \begin{tikzpicture}
    \draw (0,0) circle (1 and 1);
    \draw (0,0) circle (2 and 2);
    \draw[rotate=30] (0,0) circle (1.75 and 1.25);
    \draw[latex-latex] (0,0)--(1,0) node[pos=0.5,above]{$\Omega$};
    \draw[-] (1.1,-0.6)--(2,-1) node[right] {$\mathcal{L}_V(\alpha)$};
    \node at (-1.7,1.7) {$\mathcal{B}_{\delta^\prime}$};
  \end{tikzpicture}
  \caption{A domain of attraction and the level surface of $V$}
  \label{fig:geometric}
\end{figure}

To conclude the proof, we need to certify that $x(t_k) \in \mathcal{B}_{\delta^\prime}$ for all $k \in \bbZ_+$.
Since $\Omega$ can be arbitrarily small by restricting $\varepsilon$ to be small enough, we can choose $\alpha > 0$ for which the level surface $\mathcal{L}_V(\alpha) := \{x \in \bbR^n : V(x) = \alpha\}$ is positioned as illustrated in Fig.~\ref{fig:geometric}, where the relation $\mathcal{B}_\Omega \subset \mathcal{L}_V(\alpha) \subset \mathcal{B}_{\delta^\prime}$ holds.
In what follows, we assume that $x(t_0)$ lies inside $\mathcal{L}_V(\alpha)$.
It is clear that $x(t_k)$ remains inside $\mathcal{L}_V(\alpha)$ until $x(t_k)$ enters $\mathcal{B}_\Omega$.
Thus, we now assume that $\|x(t_k)\| \le \Omega$ for some $k \in \bbZ_+$.
In this case,
\begin{align*}
  \|x(t_{k+1})\| \le \sqrt{\frac{c_2}{c_1}} \e^{-c_3 \ell} \Omega + K \varepsilon (1 + \ell) \e^{L \ell}.
\end{align*}
Hence, for sufficiently small $\varepsilon$, it holds that $x(t_{k+1})$ is inside $\mathcal{L}_V(\alpha)$.
Therefore, we have shown that $\|x(t_k)\| < \delta^\prime$ for all $k \in \bbZ_+$ whenever the initial state lies in a small neighborhood of the origin.
The proof is now complete. \hfill \IEEEQED

\subsection{Proof of Lemma~\ref{lem:integral-bound}} \label{app:integral-bound}

From \eqref{eq:average-Jacobian}, we have
\begin{align}
  &\int_\tau^{\tau + T} [J(s,\varepsilon,z_0) - J_\mathrm{av}] \,\diff s \nonumber \\
  &\quad{} = \Gamma_2 \int_\tau^{\tau + T} \diag [\cos(R_4 \zeta(\sigma,0,z_0,\varepsilon))] \,\diff \sigma R_3. \label{eq:integral1}
\end{align}
The solution $\zeta(\tau,0,z_0,\varepsilon)$ can be written as
\begin{align*}
  \zeta(\tau,0,z_0,\varepsilon) &= z_0 + \eta \tau + \varepsilon \Gamma_4 \int_0^\tau \sin(R_4 \zeta(\sigma,0,z_0,\varepsilon)) \,\diff s.
\end{align*}
For notational simplicity, in what follows, we omit the argument $(0,z_0)$ in $\zeta$.

First, we recall that
\begin{align*}
  \cos(R_4 \zeta(\sigma,\varepsilon)) = \frac{\e^{j R_4 \zeta(\sigma,\varepsilon)} + \e^{-j R_4 \zeta(\sigma,\varepsilon)}}{2},
\end{align*}
where the exponential function is taken elementwise; for example, $\e^v = [\e^{v_1}\,\cdots\,\e^{v_n}]^\trans$ for $v = [v_1\;\cdots\;v_n]^\trans$.
Since
\begin{align*}
  \e^{\pm j R_4 \zeta(\sigma,\varepsilon)} = \e^{\pm j R_4 (z_0 + \eta \sigma)} \odot \e^{\pm j \varepsilon R_4 \Gamma_4 \int_0^\sigma \sin(R_4 \zeta(\rho,\varepsilon)) \,\diff \rho},
\end{align*}
where $\odot$ denotes the Hadamard (elementwise) product, the integral in \eqref{eq:integral1} can be calculated from
\begin{align*}
  &\int_\tau^{\tau + T} \e^{\pm j R_4 (z_0 + \eta \sigma)} \odot \e^{\pm j \varepsilon R_4 \Gamma_4 \int_0^\sigma \sin(R_4 \zeta(\rho,\varepsilon)) \,\diff \rho} \,\diff \sigma \\
  &\quad{} = \left[ \Psi_\pm(\sigma) \odot \e^{\pm j \varepsilon R_4 \Gamma_4 \int_0^\sigma \sin(R_4 \zeta(\rho,\varepsilon)) \,\diff \rho} \right]_{\tau}^{\tau + T} \\
  &\qquad{} - \int_\tau^{\tau + T} \Psi_\pm(\sigma) \odot \frac{\partial}{\partial \sigma} \e^{\pm j \varepsilon R_4 \Gamma_4 \int_0^\sigma \sin(R_4 \zeta(\rho,\varepsilon)) \,\diff \rho} \,\diff \sigma,
\end{align*}
where $\Psi_\pm(\sigma) := \mp j [\diag(R_4 \eta)]^{-1} \e^{\pm j R_4 (z_0 + \eta \sigma)}$.
Also,
\begin{align*}
  &\frac{\partial}{\partial \sigma} \e^{\pm j \varepsilon R_4 \Gamma_4 \int_0^\sigma \sin(R_4 \zeta(\rho,\varepsilon)) \,\diff \rho} \\
  &\quad{} = \pm j \varepsilon R_4 \Gamma_4 \sin(R_4 \zeta(\sigma,\varepsilon)) \odot \e^{\pm j \varepsilon R_4 \Gamma_4 \int_0^\sigma \sin(R_4 \zeta(\rho,\varepsilon)) \,\diff \rho}.
\end{align*}

Here, we note that each row of $R_4$ consists of single $1$ or $-1$ and zeros and that each entry of $\eta$ is greater than or equal to $1$ and at least one entry is exactly $1$.
Thus, we can observe that $\|\Psi_\pm(\sigma)\|_\infty = 1$ for all $\sigma \in \bbR_+$.
It follows that
\begin{align*}
  \left\| \int_\tau^{\tau + T} \e^{\pm j R_4 \zeta(\sigma,\varepsilon)} \,\diff \sigma \right\|_\infty \le 2 + \varepsilon T \|R_4 \Gamma_4\|_\infty.
\end{align*}
From \eqref{eq:integral1}, we obtain
\begin{align*}
  &\left\| \int_\tau^{\tau + T} [J(\sigma,\varepsilon,z_0) - J_\mathrm{av}] \,\diff \sigma \right\|_\infty \\
  &\quad{} \le \|\Gamma_2\|_\infty \|R_3\|_\infty (2 + \varepsilon T \|R_4 \Gamma_2\|_\infty).
\end{align*}
Be defining $K := \|\Gamma_2\|_\infty \|R_3\|_\infty \max\{2,\|R_4 \Gamma_2\|_\infty\}$,
\begin{align*}
  \left\| \frac{1}{T} \int_\tau^{\tau + T} [J(s,\varepsilon) - J_\mathrm{av}] \,\diff s \right\|_\infty \le K \left( \frac{1}{T} + \varepsilon \right).
\end{align*}
Clearly, $K \to 0$ as $\|W_\mathrm{inter}\|_\infty \to 0$.
Therefore, we conclude the proof. \hfill \IEEEQED

\subsection{Proof of Lemma~\ref{lem:integral-bound2}} \label{app:integral-bound2}

The solution $\zeta(\tau,x_0,z_0,\varepsilon)$ is given by
\begin{align}
  \zeta(\tau,x_0,z_0,\varepsilon) = z_0 + \eta \tau + \varepsilon \int_0^\tau \chi(\sigma,\varepsilon) \,\diff \sigma, \label{eq:z-solution}
\end{align}
where
\begin{align*}
  \chi(\sigma,\varepsilon) &:= \Gamma_3 \sin(R_1 \xi(\sigma,x_0,z_0,\varepsilon)) \\
  &\quad{} + \Gamma_4 \sin(R_3 \xi(\sigma,x_0,z_0,\varepsilon) + R_4 \zeta(\sigma,x_0,z_0,\varepsilon)).
\end{align*}
In what follows, we omit the argument $(x_0,z_0)$ in $\xi$ and $\zeta$.

Since
\begin{align*}
  &\int_\tau^{\tau + T} [f(x,\zeta(\sigma,\varepsilon)) - f_\mathrm{av}(x)] \,\diff \sigma \\
  &\quad{} = \Gamma_2 \int_\tau^{\tau + T} \sin(R_3 x + R_4 \zeta(\sigma,\varepsilon)) \,\diff \sigma,
\end{align*}
we need to calculate the integral of
\begin{align*}
  &\sin(R_3 x + R_4 \zeta(\sigma,\varepsilon)) \\
  &\quad{} = \frac{\e^{j (R_3 x + R_4 \zeta(\sigma,\varepsilon))} - \e^{-j (R_3 x + R_4 \zeta(\sigma,\varepsilon))}}{2j}.
\end{align*}
We note that $x$ in the above equation is fixed and is not the solution $\xi$.
Substituting \eqref{eq:z-solution} into the exponential functions, we obtain
\begin{align*}
  &\int_\tau^{\tau + T} \e^{\pm j (R_3 x + R_4 \zeta(\sigma,\varepsilon))} \,\diff \sigma \\
  &\quad{} = \e^{\pm j R_3 x} \odot \int_\tau^{\tau + T} \e^{\pm j R_4 (z_0 + \eta \sigma)} \odot \e^{\pm j \varepsilon R_4 \int_0^\sigma \chi(\rho,\varepsilon) \,\diff \rho} \,\diff \sigma.
\end{align*}
The integration by parts yields
\begin{align*}
  &\int_\tau^{\tau + T} \e^{\pm j R_4 (z_0 + \eta \sigma)} \odot \e^{\pm j \varepsilon R_4 \int_0^\sigma \chi(\rho,\varepsilon) \,\diff \rho} \,\diff \sigma \\
  &\quad{} = \left[ \Psi_\pm(\sigma) \odot \e^{\pm j \varepsilon R_4 \int_0^\sigma \chi(\rho,\varepsilon) \,\diff \rho} \right]_{\tau}^{\tau + T} \\
  &\qquad{} - \int_\tau^{\tau + T} \Psi_\pm(\sigma) \odot \frac{\partial}{\partial \sigma} \e^{\pm j \varepsilon R_4 \int_0^\sigma \chi(\rho,\varepsilon) \,\diff \rho} \,\diff \sigma,
\end{align*}
where $\Psi_\pm(\sigma) := \mp j \diag(R_4 \eta)^{-1} \e^{\pm j R_4 (z_0 + \eta \sigma)}$.
Note that
\begin{align*}
  \frac{\diff}{\diff \sigma} \e^{\pm j \varepsilon R_4 \int_0^\sigma \chi(\rho) \,\diff \rho} = \pm j \varepsilon R_4 \chi(\sigma) \odot \e^{\pm j \varepsilon R_4 \int_0^\sigma \chi(\rho) \,\diff \rho}.
\end{align*}

Since $\|\Psi_\pm(\sigma)\|_\infty = 1$ for all $\sigma \in \bbR_+$, we have
\begin{align*}
  &\left\| \int_\tau^{\tau + T} \e^{\pm j (R_3 x + R_4 \zeta(\sigma))} \,\diff \sigma \right\|_\infty \\
  &\quad{} \le 2 + \varepsilon \int_\tau^{\tau + T} \|R_4 \chi(\sigma)\|_\infty \,\diff \sigma.
\end{align*}
From the definition of $\chi$, $\|R_4 \chi(\sigma)\|_\infty = \|R_4\|_\infty (\|\Gamma_3\|_\infty + \|\Gamma_4\|_\infty)$.
Hence,
\begin{align*}
  &\left\| \int_\tau^{\tau + T} [f(x,\zeta(\sigma,\varepsilon)) - f_\mathrm{av}(x)] \,\diff \sigma \right\|_\infty \\
  &\quad{} \le \|\Gamma_2\|_\infty [2 + \varepsilon T \|R_4\|_\infty (\|\Gamma_3\|_\infty + \|\Gamma_4\|_\infty)].
\end{align*}
By defining
\begin{align*}
  K := \|\Gamma_2\|_\infty \max \{2, \|R_4\|_\infty (\|\Gamma_3\|_\infty + \|\Gamma_4\|_\infty)\},
\end{align*}
we conclude that
\begin{align*}
  &\left\| \frac{1}{T} \int_\tau^{\tau + T} [f(x,\zeta(\sigma,\varepsilon)) - f_\mathrm{av}(x)] \,\diff \sigma \right\|_\infty \\
  &\quad{} \le K \left( \frac{1}{T} + \varepsilon \right).
\end{align*}
Because $K \to 0$ as $\|W_\mathrm{inter}\| \to 0$, the proof is complete. \hfill \IEEEQED


\bibliographystyle{unsrt}
\bibliography{mybib}

\begin{thebibliography}{10}

\bibitem{Arenas_PhysRep2008}
A.~Arenas, A.~D{\'i}az-Guilera, J.~Kurths, Y.~Moreno, and C.~Zhou.
\newblock Synchronization in complex networks.
\newblock {\em Phys. Rep.}, 469(3):93--153, 2008.

\bibitem{Motter_NatPhys2013}
A.~E. Motter, S.~A. Myers, M.~Anghel, and T.~Nishikawa.
\newblock Spontaneous synchrony in power-grid networks.
\newblock {\em Nat. Phys.}, 9(3):191--197, 2013.

\bibitem{Francis_book2016}
B.~A. Francis and M.~Maggiore.
\newblock {\em Flocking and Rendezvous in Distributed Robotics}.
\newblock Springer, 2016.

\bibitem{Glass_nature2001}
L.~Glass.
\newblock Synchronization and rhythmic processes in physiology.
\newblock {\em Nature}, 410:277--284, 2001.

\bibitem{Buzsaki_science2004}
G.~Buzs{\'a}ki and A.~Draguhn.
\newblock Neuronal oscillations in cortical networks.
\newblock {\em Science}, 304(5679):1926--1929, 2004.

\bibitem{Belykh_Chaos2008}
V.~N. Belykh, G.~V. Osipov, V.~S. Petrov, J.~A.~K. Suykens, and J.~Vandewalle.
\newblock Cluster synchronization in oscillatory networks.
\newblock {\em Chaos}, 18(3):037106, 2008.

\bibitem{Lu_Chaos2010}
W.~Lu, B.~Liu, and T.~Chen.
\newblock Cluster synchronization in networks of coupled nonidentical dynamical
  systems.
\newblock {\em Chaos}, 20(1):013120, 2010.

\bibitem{Nicosia_PRL2013}
V.~Nicosia, M.~Valencia, M.~Chavez, A.~D\'{\i}az-Guilera, and V.~Latora.
\newblock Remote synchronization reveals network symmetries and functional
  modules.
\newblock {\em Phys. Rev. Lett.}, 110(17):174102, 2013.

\bibitem{Cho_PRL2017}
Y.~S. Cho, T.~Nishikawa, and A.~E. Motter.
\newblock Stable chimeras and independently synchronizable clusters.
\newblock {\em Phys. Rev. Lett.}, 119(8):084101, 2017.

\bibitem{Stam_2012}
C.~J. Stam and E.~C.~W. {van Straaten}.
\newblock The organization of physiological brain networks.
\newblock {\em Clinical Neurophysiology}, 123(6):1067--1087, 2012.

\bibitem{Hegselmann_2002}
R.~Hegselmann and U.~Krause.
\newblock Opinion dynamics and bounded confidence: Models, analysis and
  simulation.
\newblock {\em Journal of Artificial Societies and Social Simulation}, 5(3),
  2002.

\bibitem{Bassett_2009}
D.~S. Bassett and E.~T. Bullmore.
\newblock Human brain networks in health and disease.
\newblock {\em Curr. Opin. Neurol.}, 22(4):340--347, 2009.

\bibitem{Pham2007}
Q.-C. Pham and J.-J.~E. Slotine.
\newblock Stable concurrent synchronization in dynamic system networks.
\newblock {\em Neural Netw.}, 20(1):62--77, 2007.

\bibitem{Pogromsky_2008}
A.~Y. Pogromsky.
\newblock A partial synchronization theorem.
\newblock {\em Chaos}, 18:037107, 2008.

\bibitem{Wu2009}
W.~Wu, W.~Zhou, and T.~Chen.
\newblock Cluster synchronization of linearly coupled complex networks under
  pinning control.
\newblock {\em IEEE Trans. Circuits Syst. I, Reg. Papers}, 56(4):829--839,
  2009.

\bibitem{Lu_PLA2009}
X.~B. Lu and B.~Z. Qin.
\newblock Adaptive cluster synchronization in complex dynamical networks.
\newblock {\em Phys. Lett. A}, 373(40):3650--3658, 2009.

\bibitem{Liu2011}
X.~Liu and T.~Chen.
\newblock Cluster synchronization in directed networks via intermittent pinning
  control.
\newblock {\em IEEE Trans. Neural Netw.}, 22(7):1009--1020, 2011.

\bibitem{Xia_2011}
W.~Xia and M.~Cao.
\newblock Clustering in diffusively coupled networks.
\newblock {\em Automatica}, 47(11):2395--2405, 2011.

\bibitem{Kuramoto_book1984}
Y.~Kuramoto.
\newblock {\em Chemical Oscillations, Waves, and Turbulence}.
\newblock Springer, 1984.

\bibitem{Acebron_RMP2005}
J.~A. Acebr\'{o}n, L.~L. Bonilla, C.~J. {P\'{e}rez Vicente}, F.~Ritort, and
  R.~Spigler.
\newblock The {Kuramoto} model: A simple paradigm for synchronization
  phenomena.
\newblock {\em Rev. Mod. Phys.}, 77(1):137--185, 2005.

\bibitem{Jadbabaie_ACC2004}
A.~Jadbabaie, N.~Motee, and M.~Barahona.
\newblock On the stability of the {Kuramoto} model of coupled nonlinear
  oscillators.
\newblock In {\em Proc. Am. Control Conf.}, pages 4296--4301, 2004.

\bibitem{Dorfler_Automatica2014}
F.~D{\"o}rfler and F.~Bullo.
\newblock Synchronization in complex networks of phase oscillators: A survey.
\newblock {\em Automatica}, 50(6):1539--1564, 2014.

\bibitem{Wu-Li_2020}
J.~Wu and X.~Li.
\newblock Collective synchronization of {Kuramoto}-oscillator networks.
\newblock {\em IEEE Circuits and Systems Magazine}, 20(3):46--67, 2020.

\bibitem{Schaub_Chaos2016}
M.~T. Schaub, N.~O'Clery, Y.~N. Billeh, J.-C. Delvenne, R.~Lambiotte, and
  M.~Barahona.
\newblock Graph partitions and cluster synchronization in networks of
  oscillators.
\newblock {\em Chaos}, 26(9):094821, 2016.

\bibitem{Menara_TCNS2020}
T.~Menara, G.~Baggio, D.~S. Bassett, and F.~Pasqualetti.
\newblock Stability conditions for cluster synchronization in networks of
  heterogeneous {Kuramoto} oscillators.
\newblock {\em IEEE Trans. Control Netw. Syst.}, 7(1):302--314, 2020.

\bibitem{Menara_ACC2019}
T.~Menara, G.~Baggio, D.~S. Bassett, and F.~Pasqualetti.
\newblock Exact and approximate stability conditions for cluster
  synchronization of {Kuramoto} oscillators.
\newblock In {\em Proc. Am. Control Conf.}, pages 205--210, 2019.

\bibitem{Qin_TAC2021}
Y.~Qin, Y.~Kawano, O.~Portoles, and M.~Cao.
\newblock Partial phase cohesiveness in networks of networks of {Kuramoto}
  oscillators.
\newblock {\em IEEE Trans. Autom. Control}, 66(12):6100--6107, 2021.

\bibitem{Tiberi_CDC2017}
L.~Tiberi, C.~Favaretto, M.~Innocenti, D.~S. Bassett, and F.~Pasqualetti.
\newblock Synchronization patterns in networks of {Kuramoto} oscillators: A
  geometric approach for analysis and control.
\newblock In {\em Proc. IEEE Conf. on Decision and Control}, pages 481--486,
  2017.

\bibitem{Menara_CDC2019}
T.~Menara, G.~Baggio, D.~S. Bassett, and F.~Pasqualetti.
\newblock A framework to control functional connectivity in the human brain.
\newblock In {\em Proc. IEEE Conf. Decis. Control}, pages 4697--4704, 2019.

\bibitem{Sanders-Verhulst-Murdock_book2007}
J.~A. Sanders, F.~Verhulst, and J.~Murdock.
\newblock {\em Averaging Methods in Nonlinear Dynamical Systems}.
\newblock Springer, 2nd edition, 2007.

\bibitem{Hapaev_book1993}
M.~M. Hapaev.
\newblock {\em Averaging in Stability Theory}.
\newblock Springer, 1993.

\bibitem{Hale_book1980}
J.~K. Hale.
\newblock {\em Ordinary Differential Equations}.
\newblock Robert E. Krieger, 2nd edition, 1980.

\bibitem{Khalil_book2002}
H.~K. Khalil.
\newblock {\em Nonlinear Systems}.
\newblock Prentice Hall, 3rd edition, 2002.

\bibitem{Aeyels_TAC1998}
D.~Aeyels and J.~Peuteman.
\newblock A new asymptotic stability criterion for nonlinear time-variant
  differential equations.
\newblock {\em IEEE Trans. Autom. Control}, 43(7):968--971, 1998.

\bibitem{Aeyels_TAC1999}
D.~Aeyels and J.~Peuteman.
\newblock On exponential stability of nonlinear time-varying differential
  equations.
\newblock {\em Automatica}, 35(6):1091--1100, 1999.

\bibitem{Michel-Hou-Liu_book2015}
A.~N. Michel, L.~Hou, and D.~Liu.
\newblock {\em Stability of Dynamical Systems}.
\newblock Birkh\"{a}user, 2nd edition, 2015.

\bibitem{Vorotnikov_2005}
V.~I. Vorotnikov.
\newblock Partial stability and control: The state-of-the-art and development
  prospects.
\newblock {\em Autom. Remote Control}, 66(4):511--561, 2005.

\bibitem{Kato_ECC2021}
R.~Kato and H.~Ishii.
\newblock Averaging and cluster synchronization of {Kuramoto} oscillators.
\newblock In {\em Proc. Eur. Control Conf.}, pages 1497--1502, 2021.

\bibitem{Godsil-Royle_book2001}
C.~Godsil and G.~Royle.
\newblock {\em Algebraic Graph Theory}.
\newblock Springer, 2001.

\bibitem{Menara_CSL2022}
T.~Menara, Y.~Qin, D.~S. Bassett, and F.~Pasqualetti.
\newblock Relay interactions enable remote synchronization in networks of phase
  oscillators.
\newblock {\em IEEE Control Syst. Lett.}, 6:500--505, 2022.

\bibitem{Cardoso2007}
D.~M. Cardoso, C.~Delorme, and P.~Rama.
\newblock {Laplacian} eigenvectors and eigenvalues and almost equitable
  partitions.
\newblock {\em Eur. J. Comb.}, 28(3):665--673, 2007.

\bibitem{Gambuzza_Automatica2019}
L.~V. Gambuzza and M.~Frasca.
\newblock A criterion for stability of cluster synchronization in networks with
  external equitable partitions.
\newblock {\em Automatica}, 100:212--218, 2019.

\bibitem{Pecora2014}
L.~M. Pecora, F.~Sorrentino, A.~M. Hagerstrom, T.~E. Murphy, and R.~Roy.
\newblock Cluster synchronization and isolated desynchronization in complex
  networks with symmetries.
\newblock {\em Nat. Commun.}, 5:4079, 2014.

\bibitem{Sorrentino2016}
F.~Sorrentino, L.~M. Pecora, A.~M. Hagerstrom, T.~E. Murphy, and R.~Roy.
\newblock Complete characterization of the stability of cluster synchronization
  in complex dynamical networks.
\newblock {\em Sci. Adv.}, 2(4):e1501737, 2016.

\bibitem{O'Clery_PRE2013}
N.~O'Clery, Y.~Yuan, G.-B. Stan, and M.~Barahona.
\newblock Observability and coarse graining of consensus dynamics through the
  external equitable partition.
\newblock {\em Phys. Rev. E}, 88(4):042805, 2013.

\bibitem{Zhang_TAC2014}
S.~Zhang, M.~Cao, and M.~K. Camlibel.
\newblock Upper and lower bounds for controllable subspaces of networks of
  diffusively coupled agents.
\newblock {\em IEEE Trans. Autom. Control}, 59(3):745--750, 2014.

\bibitem{Haddad-Chellaboina_book2008}
W.~M. Haddad and V.~S. Chellaboina.
\newblock {\em Nonlinear Dynamical Systems and Control: A {Lyapunov}-Based
  Approach}.
\newblock Princeton University Press, 2008.

\bibitem{Qin_TAC2021b}
Y.~Qin, Y.~Kawano, B.~D.~O. Anderson, and M.~Cao.
\newblock Partial exponential stability analysis of slow-fast systems via
  periodic averaging.
\newblock {\em IEEE Trans. Autom. Control}, 67(10):5479--5486, 2022.

\bibitem{Teel2003}
A.~R. Teel, L.~Moreau, and D.~Ne\v{s}i\'{c}.
\newblock A unified framework for input-to-state stability in systems with two
  time scales.
\newblock {\em IEEE Trans. Autom. Control}, 48(9):1526--1544, 2003.

\bibitem{Miroshnik_2002}
I.~V. Miroshnik.
\newblock Partial stability and geometric problems of nonlinear dynamics.
\newblock {\em Autom. Remote Control}, 63(11):1730--1744, 2002.

\bibitem{Hancock_2014}
E.~J. Hancock and D.~J. Hill.
\newblock Restricted partial stability and synchronization.
\newblock {\em IEEE Transactions on Circuits and Systems I: Regular Papers},
  61(11):3235--3244, 2014.

\bibitem{Kosut1987}
R.~L. Kosut, B.~D.~O. Anderson, and I.~M.~Y. Mareels.
\newblock Stability theory for adaptive systems: Method of averaging and
  persistency of excitation.
\newblock {\em IEEE Trans. Autom. Control}, 32(1):26--34, 1987.

\bibitem{Stilwell_SIADS2006}
D.~J. Stilwell, E.~M. Bollt, and D.~G. Roberson.
\newblock Sufficient conditions for fast switching synchronization in
  time-varying network topologies.
\newblock {\em SIAM J. Appl. Dyn. Syst.}, 5(1):140--156, 2006.

\bibitem{Teel-Peuteman-Aeyels_1999}
A.~R. Teel, J.~Peuteman, and D.~Aeyels.
\newblock Semi-global practical asymptotic stability and averaging.
\newblock {\em Systems \& Control Letters}, 37:329--334, 1999.

\bibitem{Rugh_book1996}
W.~J. Rugh.
\newblock {\em Linear System Theory}.
\newblock Prentice Hall, 2nd edition, 1996.

\bibitem{Adrianova_book1995}
L.~Y. Adrianova.
\newblock {\em Introduction to Linear Systems of Differential Equations}.
\newblock Translations of Mathematical Monographs. AMS, 1995.

\bibitem{Altafini_TAC2013}
C.~Altafini.
\newblock Consensus problems on networks with antagonistic interactions.
\newblock {\em IEEE Trans. Autom. Control}, 58(4):935--946, 2013.

\bibitem{Proskurnikov_TAC2016}
A.~V. Proskurnikov, A.~S. Matveev, and M.~Cao.
\newblock Opinion dynamics in social networks with hostile camps: Consensus vs.
  polarization.
\newblock {\em IEEE Trans. Autom. Control}, 61(6):1524--1536, 2016.

\bibitem{Delabays_SIADS2019}
R.~Delabays, P.~Jacquod, and F.~D\"{o}rfler.
\newblock The {Kuramoto} model on oriented and signed graphs.
\newblock {\em SIAM J. Appl. Dyn. Syst.}, 18(1):458--480, 2019.

\bibitem{Montenbruck_Automatica2015}
J.~M. Montenbruck, M.~B{\"u}rger, and F.~Allg{\"o}wer.
\newblock Practical synchronization with diffusive couplings.
\newblock {\em Automatica}, 53:235--243, 2015.

\bibitem{Lageman_CDC2016}
C.~Lageman and Z.~Sun.
\newblock Consensus on spheres: Convergence analysis and perturbation theory.
\newblock In {\em Proc. IEEE Conf. Decis. Control}, pages 19--24, 2016.

\bibitem{Cabral2011}
J.~Cabral, E.~Hugues, O.~Sporns, and G.~Deco.
\newblock Role of local network oscillations in resting-state functional
  connectivity.
\newblock {\em NeuroImage}, 57(1):130--139, 2011.

\bibitem{Sporns_book2011}
O.~Sporns.
\newblock {\em Networks of the Brain}.
\newblock MIT Press, 2011.

\bibitem{Hagmann_plosb2008}
P.~Hagmann, L.~Cammoun, X.~Gigandet, R.~Meuli, C.~J. Honey, V.~J. Wedeen, and
  O.~Sporns.
\newblock Mapping the structural core of human cerebral cortex.
\newblock {\em PLoS Biol.}, 6(7):e159, 2008.

\bibitem{Brown12}
J.~Brown, J.~Rudie, A.~Bandrowski, J.~VanHorn, and S.~Bookheimer.
\newblock The {UCLA} multimodal connectivity database: A web-based platform for
  brain connectivity matrix sharing and analysis.
\newblock {\em Front. Neuroinform.}, 6:28, 2012.

\bibitem{Honey_PNAS2009}
C.~J. Honey, O.~Sporns, L.~Cammoun, X.~Gigandet, J.~P. Thiran, R.~Meuli, and
  P.~Hagmann.
\newblock Predicting human resting-state functional connectivity from
  structural connectivity.
\newblock {\em Proc. Natl. Acad. Sci. USA}, 106(6):2035--2040, 2009.

\bibitem{Xia_ploso2013}
M.~Xia, J.~Wang, and Y.~He.
\newblock {BrainNet Viewer}: A network visualization tool for human brain
  connectomics.
\newblock {\em PLoS One}, 8(7):e68910, 2013.

\bibitem{Amari_book}
S.-I. Amari.
\newblock {\em Mathematical Theory of Nerve Nets}.
\newblock Sangyo Tosho, 1978 (in Japanese).

\bibitem{Wilson_Biophys1972}
H.~R. Wilson and J.~D. Cowan.
\newblock Excitatory and inhibitory interactions in localized populations of
  model neurons.
\newblock {\em Biophys. J.}, 12(1):1--24, 1972.

\bibitem{Sadilek_2015}
M.~Sadilek and S.~Thurner.
\newblock Physiologically motivated multiplex kuramoto model describes phase
  diagram of cortical activity.
\newblock {\em Sci. Rep.}, 5:10015, 2015.

\bibitem{Bullmore_NatRevNeurosci2009}
E.~Bullmore and O~Sporns.
\newblock Complex brain networks: Graph theoretical analysis of structural and
  functional systems.
\newblock {\em Nat. Rev. Neurosci.}, 10(3):186--198, 2009.

\bibitem{Deco_PNAS2009}
G.~Deco, V.~Jirsa, A.~R. McIntosh, O.~Sporns, and R.~K\"{o}tter.
\newblock Key role of coupling, delay, and noise in resting brain fluctuations.
\newblock {\em Proc. Natl. Acad. Sci. USA}, 106(25):10302--10307, 2009.

\bibitem{Punetha_2015}
N.~Punetha, S.~R. Ujjwal, F.~M. Atay, and R.~Ramaswamy.
\newblock Delay-induced remote synchronization in bipartite networks of phase
  oscillators.
\newblock {\em Phys. Rev. E}, 91:022922, 2015.

\bibitem{Baksalary2007}
J.~K. Baksalary and O.~M. Baksalary.
\newblock Particular formulae for the {Moore--Penrose} inverse of a columnwise
  partitioned matrix.
\newblock {\em Linear Algebra Its Appl.}, 421(1):16--23, 2007.

\end{thebibliography}

\end{document}